\documentclass[11pt]{article}

\usepackage{graphics}
\usepackage[dvips]{graphicx}
\usepackage[left]{lineno}
\usepackage{multicol}
\usepackage{float}
\usepackage{amsmath}
\usepackage{subfigure}
\usepackage{eqnarray}
\usepackage{amsfonts}
\usepackage{amssymb}
\usepackage{commath}
\usepackage{subfigure}

\newcommand{\autocov}[1]{\mathtt{cov}_{#1} }

\begin{document}


\vspace{1.2cm}

\begin{center}

{\bf  {\Large The nonlinear dynamics of a bistable energy 
harvesting system with colored noise disturbances}}
\bigskip

{\small Vinicius Gonçalves Lopes \footnote{vinicius.g.lopes@uerj.br}, 
	    Jo\~{a}o Victor Ligier Lopes Peterson \footnote{joao.peterson@uerj.br} \\ and
		Americo Cunha~Jr \footnote{americo@ime.uerj.br}}
\smallskip

{\small Universidade do Estado do Rio de Janeiro -- UERJ, Rio de Janeiro -- RJ, Brazil }

{\footnotesize Received on February **, 2019 / accepted on *****, 2019}

\end{center}

\quad
\begin{abstract}

This paper deals with the nonlinear stochastic dynamics
of a piezoelectric energy harvesting system subjected to a 
harmonic external excitation disturbed by a Gaussian colored noise.
A parametric analysis is conducted, where the effects of the standard 
deviation and the correlation time of colored noise on the system 
response are investigated. The numerical results suggests a strong 
influence of noise on the system response for higher values of correlation 
time and standard deviation, and a low (noise level independent) influence
for low values of correlation time.

\end{abstract}

\quad

\textbf{1. Introduction}
\smallskip

Energy harvesting systems have been widely 
studied in literature on the past 15 years, because
of their great appeal in many different applications: 
autonomous power supply to small 
electrical-electronic boarded equipment; 
secondary energy recovering in engineering;
Internet-Of-Things (IoF) technologies in
telecommunications; 
pacemarks in health equipment; among others.

Since the pioneering work of Cottone et al. \cite{Cottone2009},
a special attention is given to the study of nonlinear energy harvesters,
once the use of nonlinearities may enhance a lot power recovering capability 
in comparison to linear energy harvesting systems. For this reason, many recent works 
\cite{Bradai2015,Kamalinejad2015,Vijayan2015,Abdelkareem2018,Wang2018}
has devoted attention to the subject.

\pagebreak
In the case of a vibrating harvester, its energetic efficiency is strongly influenced 
by the amplitude and frequency characteristics of the external excitation 
\cite{Cunhajr_matec2018}, and the introduction of a random component in the 
excitation source, for example, may raise the amount of harvested power 
\cite{Borowiec2015,Langley2015}. Investigate in detail how the characteristics 
of a forcing noise affect the response of a energy harvester is a relevant research 
problem in several areas of engineering and applied physics.

In this context, the present work deals with the nonlinear dynamics of a 
bistable piezo-magneto-elastic energy harvesting system subjected 
to an external forcing composed of a harmonic part disturbed by a 
random noise. The effects of the correlation time between harmonic 
forcing period and the random noise signal on harvester output voltage 
are investigated considering different levels of intensity for the noise disturbance.

\quad

\textbf{2. Energy harvesting system}
\label{sec_models}
\smallskip

\begin{figure}[!h]
\centering
\includegraphics[scale=0.35]{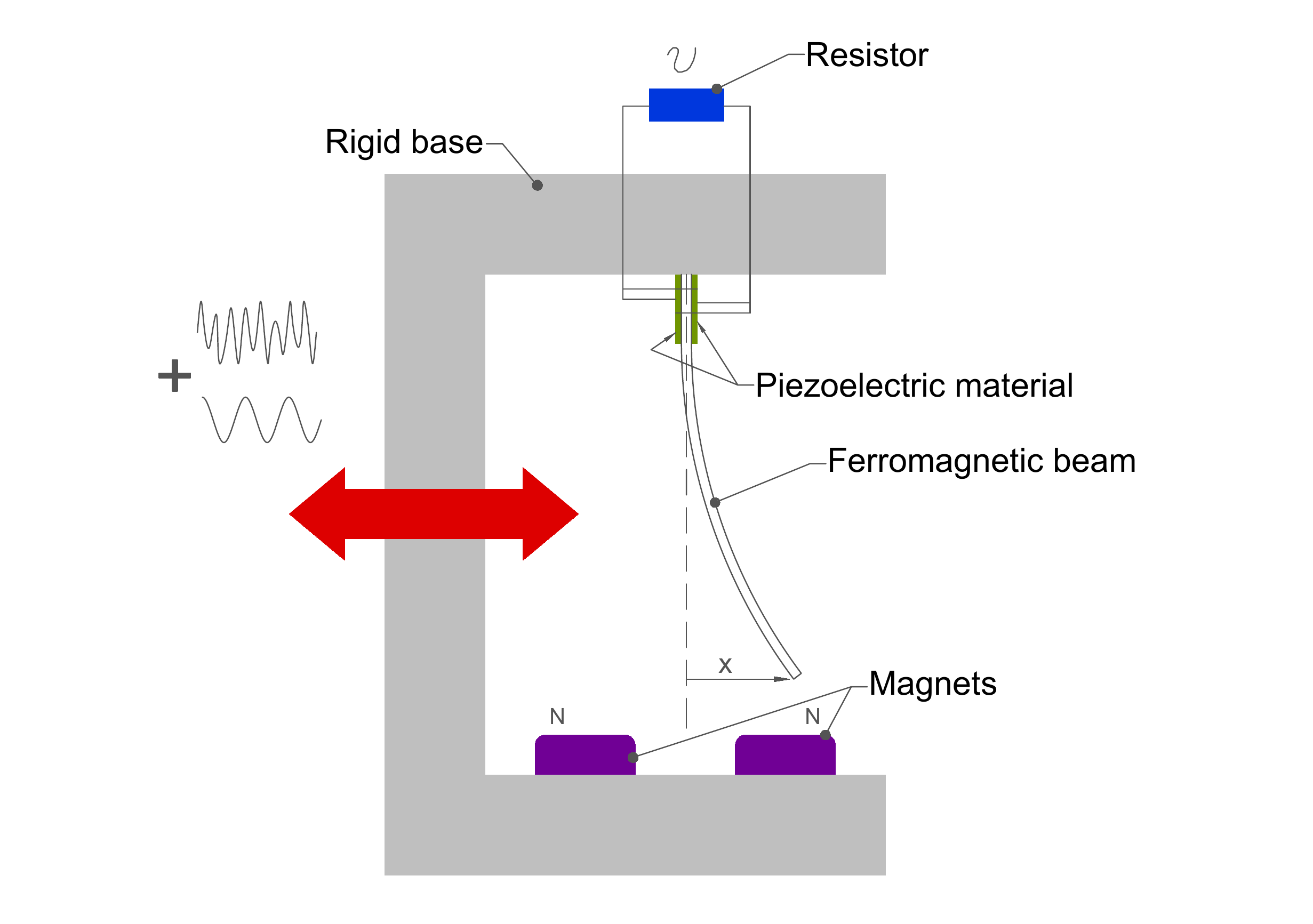}
\caption{Piezo-magneto-elastic energy harvesting system.}
\label{fig_bistable_harvesting}
\end{figure}

The energy harvesting system studied here, 
proposed by Erturk et al. \cite{Erturk2009}, is the 
piezo-magneto-elastic beam, driven by a
rigid base movement, that is illustrated in 
Figure \ref{fig_bistable_harvesting}. 

Its nonlinear stochastic dynamics evolves according to
the following system of stochastic differential equations

\begin{equation}
\ddot{X} + 2 \xi \dot{X} - \frac{1}{2} \, X \, (1-X^2) - \chi V = f \cos{\Omega t} + N_{t},
\label{eq_mechanical_model}
\end{equation}

\vspace{-0.1cm}
\begin{equation}
\dot{V} + \lambda V + \kappa \dot{X} = 0,
\label{eq_electrical_model}
\end{equation}

\noindent 
where the beam free edge displacement, its velocity
and the system output voltage are represented by the
random processes $X$, $\dot{X}$, and $V$, respectively;
$\xi$ means the mechanical damping ratio; 
$\chi$ and $\kappa$ are the piezoelectric coupling terms; 
and $\lambda$ is a reciprocal time constant;
Such parameters are dimensionless, assumed as $\xi=0.01$, 
$\chi=0.05$, $\kappa=0.5$, $\lambda=0.05$, and the 
initial conditions are adopted as being
$X_{0} = 1$, $\dot{X}_0 = 0$ and $V_{0} = 0$.

The external excitation is compound a harmonic 
force with amplitude and frequency respectively given by
$f = 0.115$ and $\Omega = 0.8$, and disturbance 
component $N_{t}$, assumed as a zero-mean Gaussian 
colored noise with covariance function

\begin{equation}
\autocov{N_{t}} (t_{1}, t_{2}) = \sigma \, \exp{ \left(- \frac{|t_{2} - t_{1}|}{\tau_{corr}} \right) },
\label{eq_covariance_noise}
\end{equation}

\noindent
where $\tau_{corr}$ means the correlation time and $\sigma$ the 
colored noise standard deviation. The parameters values are 
assumed as percentages of harmonic forcing amplitude and 
frequency, respectively.

\quad

\textbf{3. Numerical experimentation}
\label{sec_numerical_experimentation}
\smallskip

For the present analysis, the noise correlation times are set as 
$\tau_{corr} =  \eta \, \tau_{\Omega}$, where $\tau_{\Omega} = 2\pi/\Omega$ and
$\eta = \left\lbrace 1\%,~ 50\%, ~100\% \right\rbrace $. The noise standard deviation values 
are assumed as $1\%$, $25\%$ and $50\%$ of harmonic amplitude $f$.
In each case, a Monte Carlo simulation with 256 samples is performed.
The realizations of noise are generated by deterministic approximations, 
obtained by the Karhunen-Loève expansion. This method allows the 
dynamics to be integrated with the standard 4th-order Runge-Kutta method 
with automatic time step.

\begin{figure}[ht!]
\centering 
\subfigure[$\tau_{corr}/\tau_{\Omega} = 1\%$]{\includegraphics[width = 3.8cm,height = 4cm]{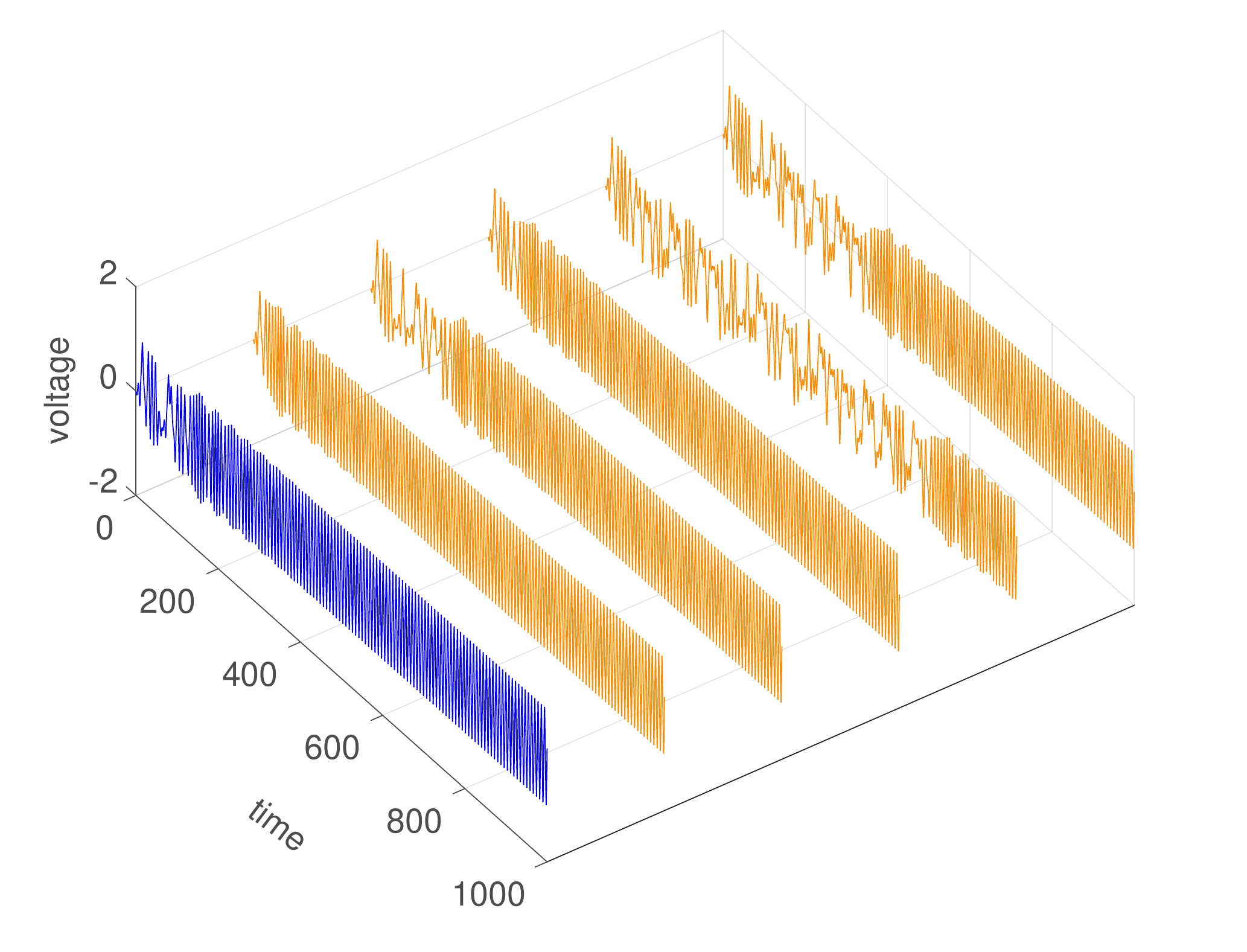}} \quad
\subfigure[$\tau_{corr}/\tau_{\Omega} = 50\%$]{\includegraphics[width = 3.8cm,height = 4cm]{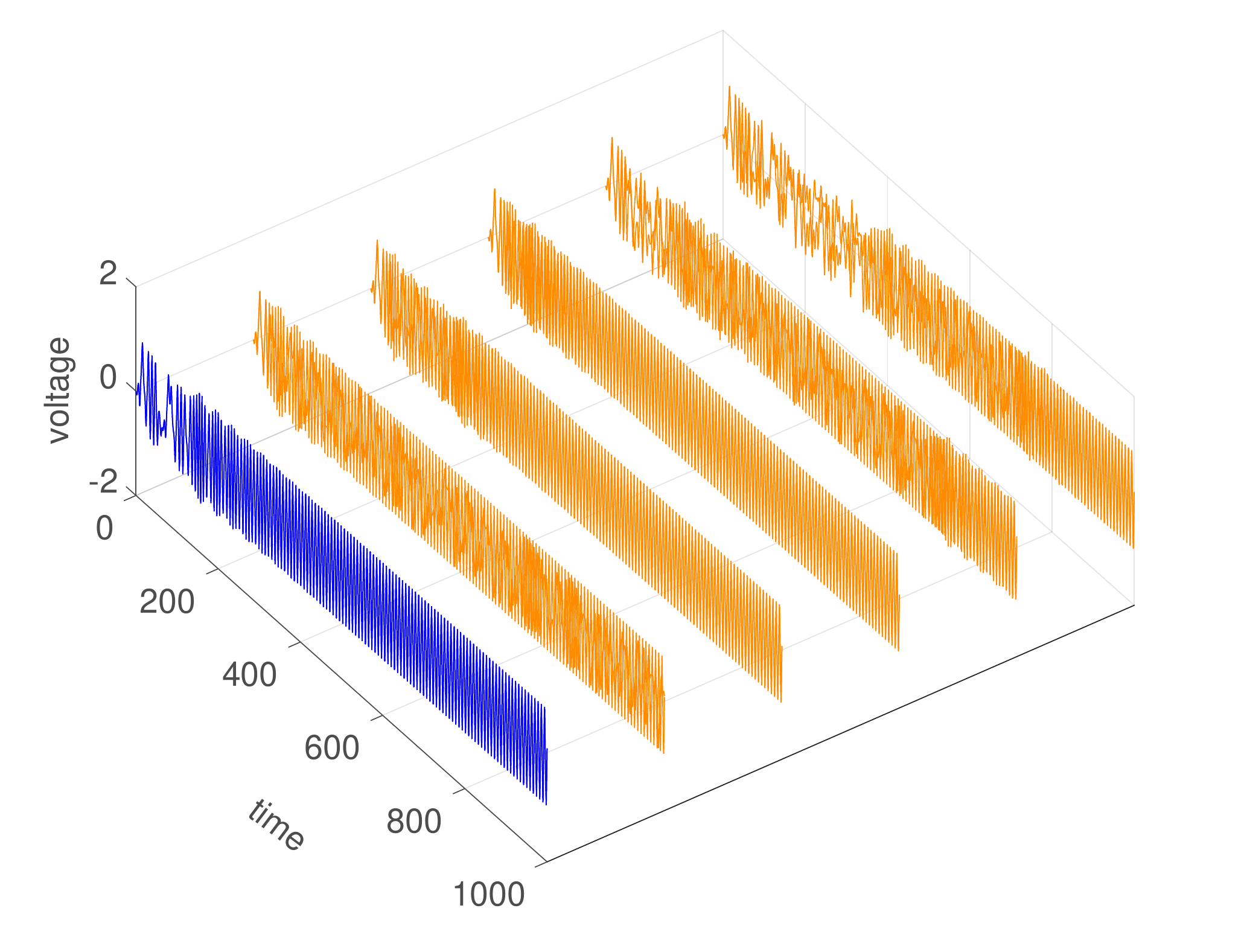}} \quad
\subfigure[$\tau_{corr}/\tau_{\Omega} = 100\%$]{\includegraphics[width = 3.8cm,height = 4cm]{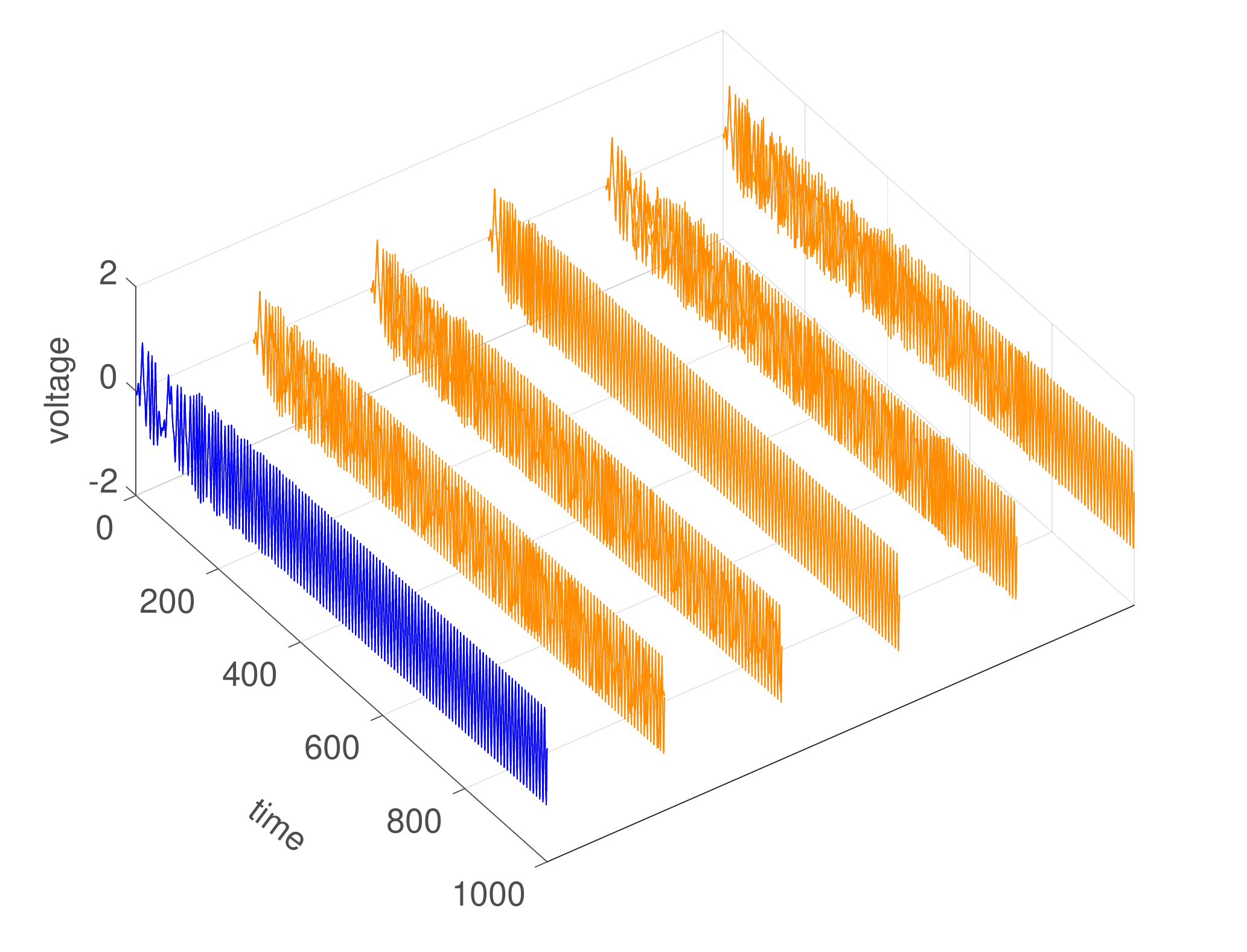}}

\subfigure[$\tau_{corr}/\tau_{\Omega} = 1\%$]{\includegraphics[width = 3.8cm,height = 4cm]{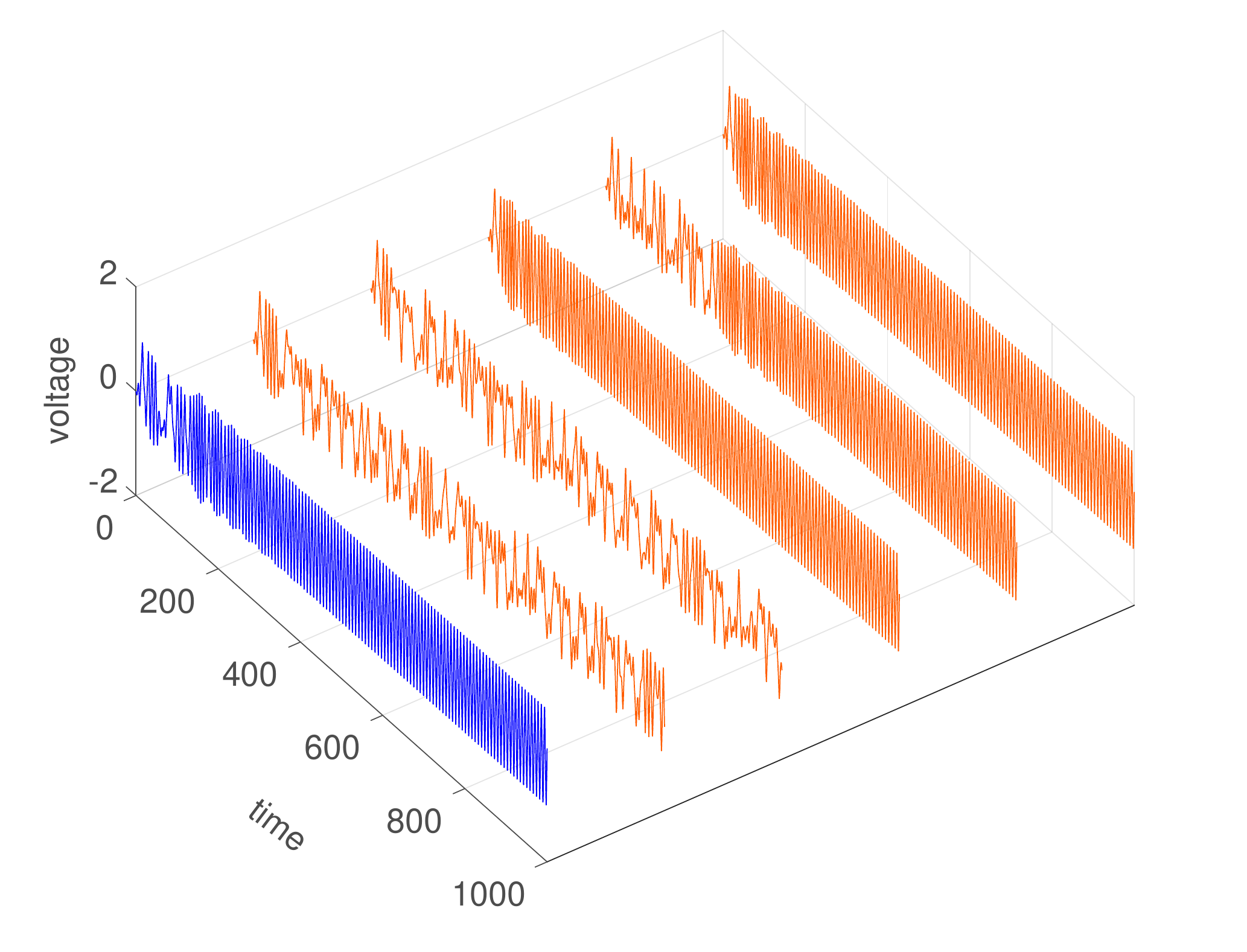}}\quad
\subfigure[$\tau_{corr}/\tau_{\Omega} = 50\%$]{\includegraphics[width = 3.8cm,height = 4cm]{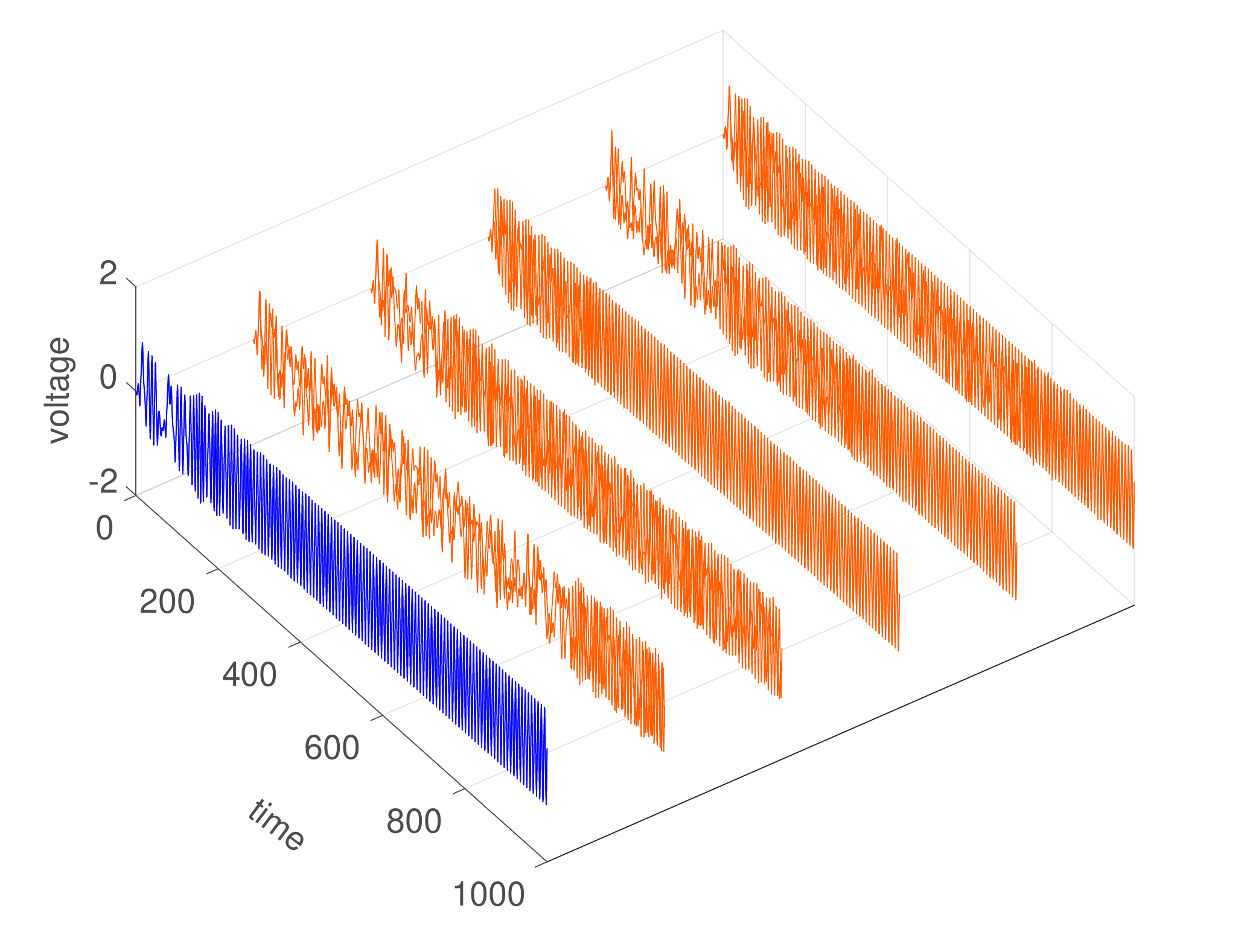}} \quad
\subfigure[$\tau_{corr}/\tau_{\Omega} = 100\%$]{\includegraphics[width = 3.8cm,height = 4cm]{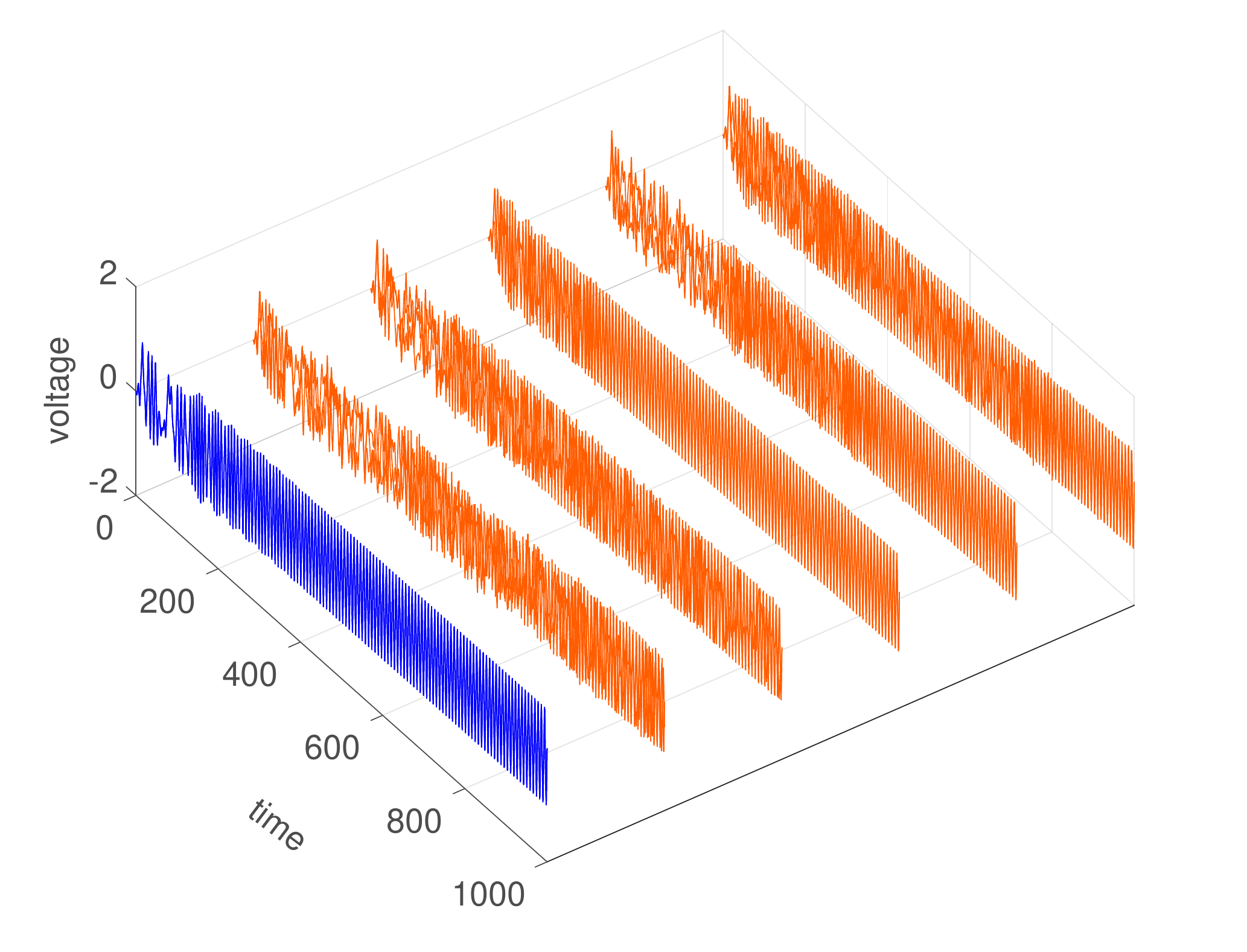}}

\subfigure[$\tau_{corr}/\tau_{\Omega} = 1\%$]{\includegraphics[width = 3.8cm,height = 4cm]{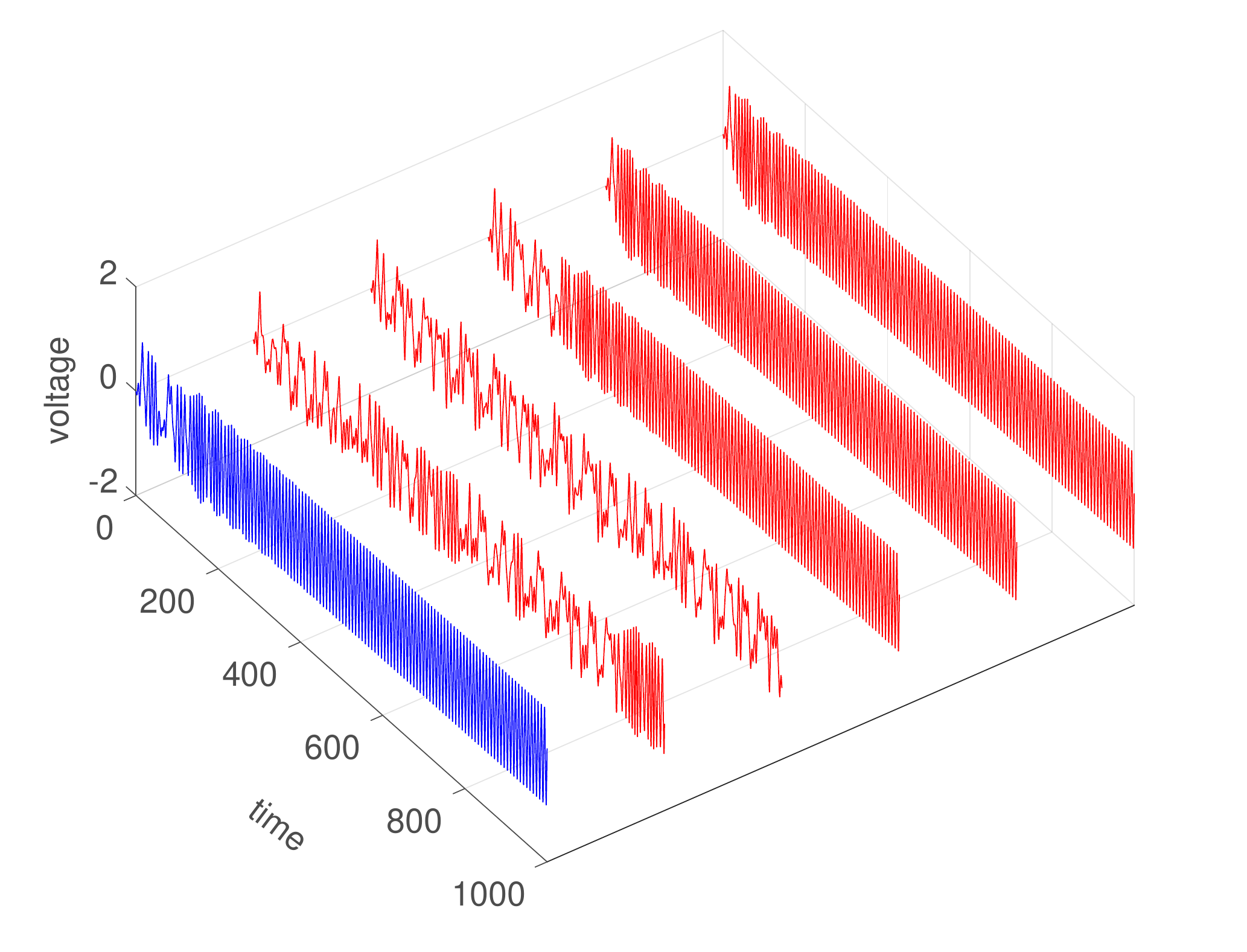}}
\quad
\subfigure[$\tau_{corr}/\tau_{\Omega} = 50\%$]{\includegraphics[width = 3.8cm,height = 4cm]{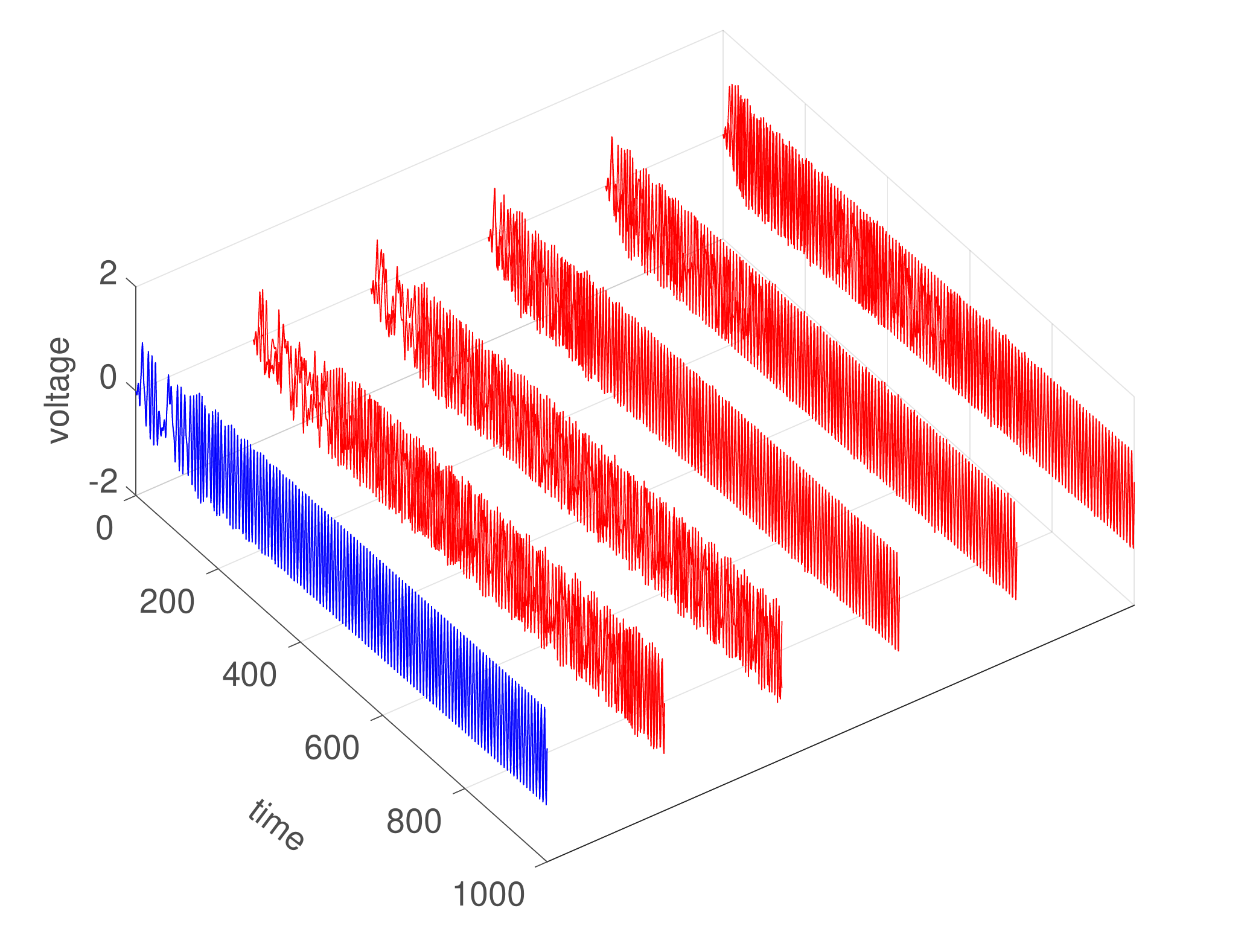}} 
\quad
\subfigure[$\tau_{corr}/\tau_{\Omega} = 100\%$]{\includegraphics[width = 3.8cm,height = 4cm]{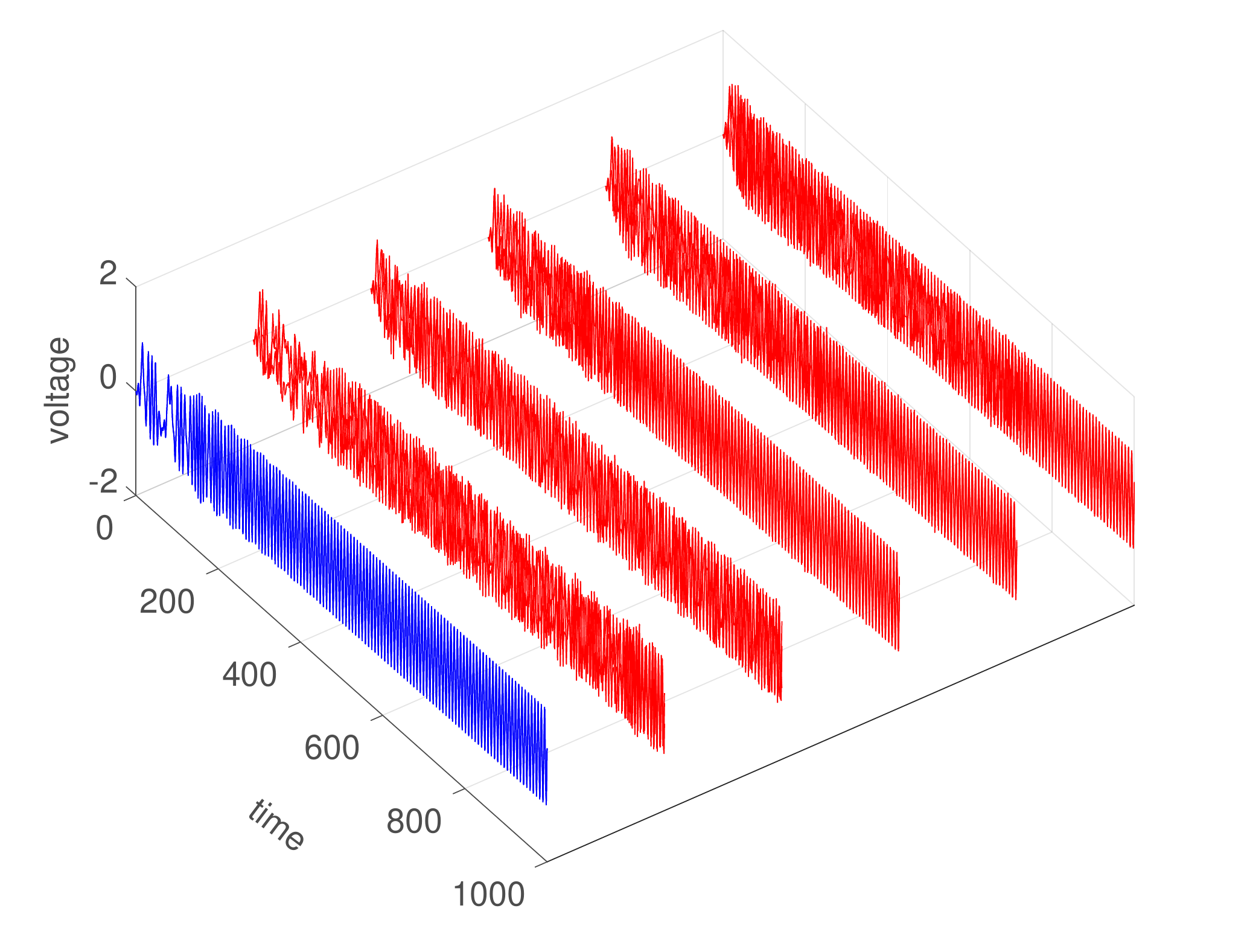}}

\caption{Typical samples of voltage time series for $\sigma/f = 1\%$ (top), $\sigma/f = 25\%$ (middle) 
and $\sigma/f = 50\%$ (bottom), and different values of correlation time. 
On blue, the deterministic case, considering harmonic forcing.}
\label{fig_time_series_deltaN}
\end{figure}

\begin{figure}[ht!]
\centering 
\subfigure[$\tau_{corr}/\tau_{\Omega} = 1\%$]{\includegraphics[scale=0.2]{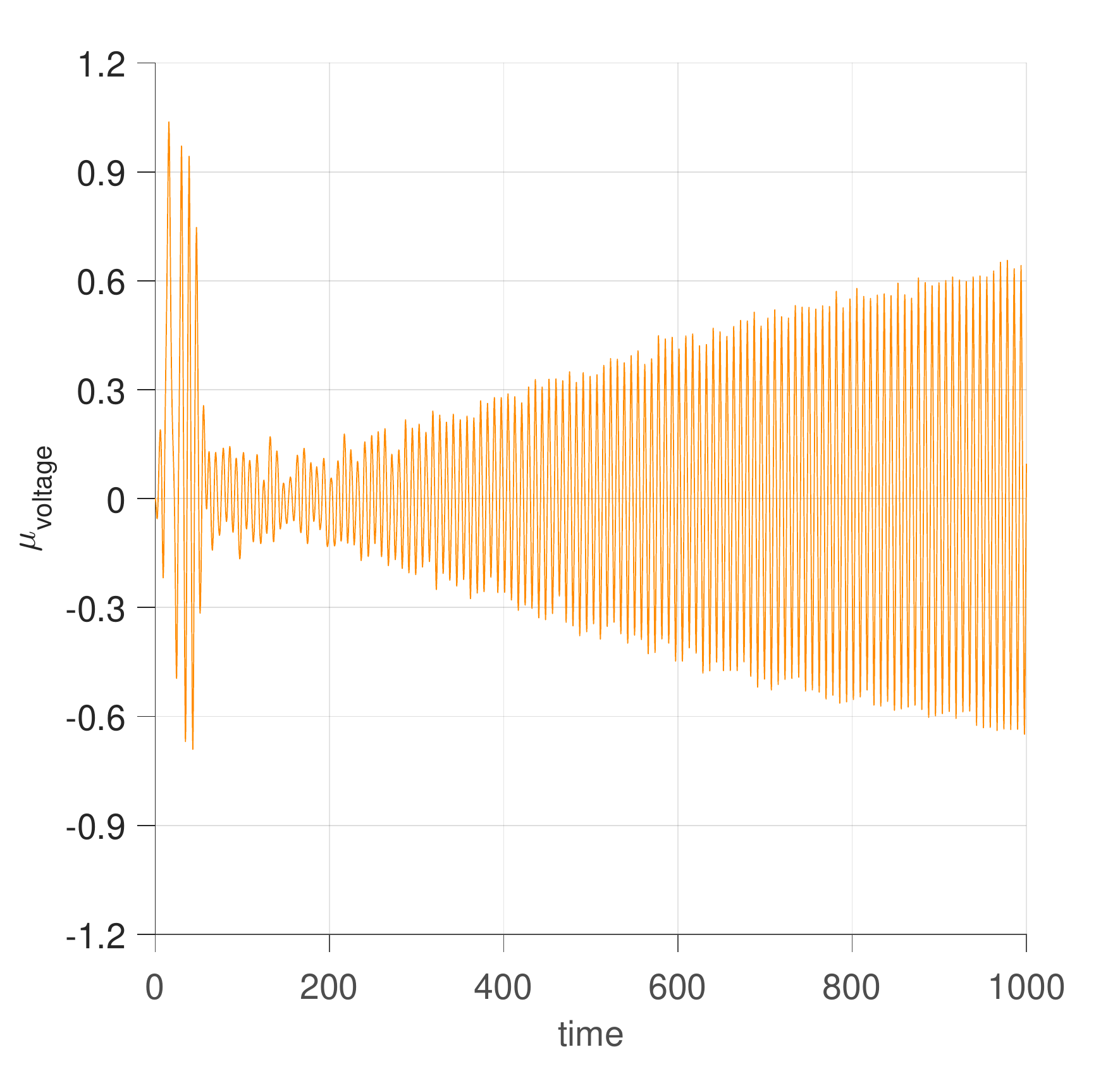}}
\subfigure[$\tau_{corr}/\tau_{\Omega} = 50\%$]{\includegraphics[scale=0.2]{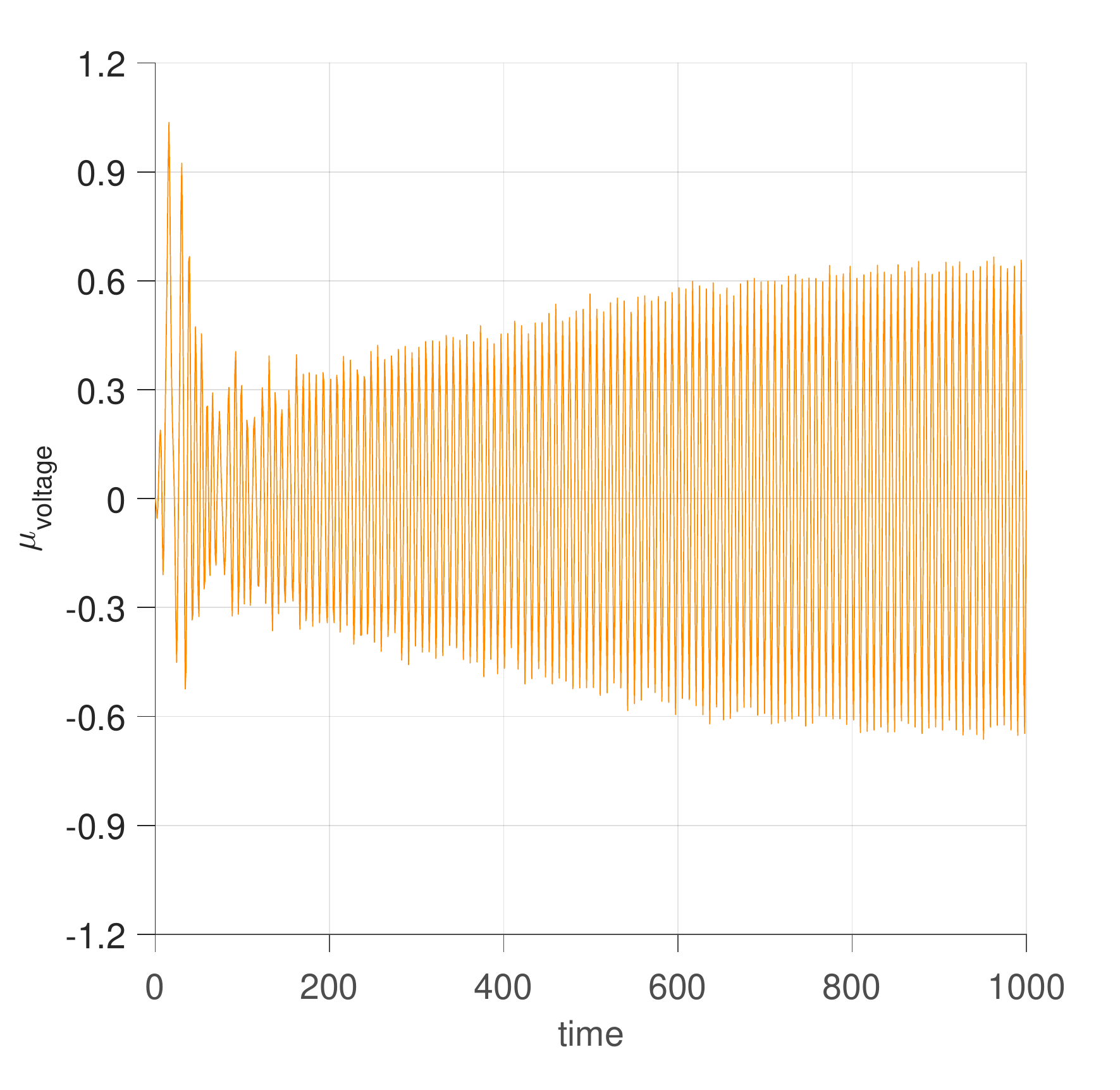}}
\subfigure[$\tau_{corr}/\tau_{\Omega} = 100\%$]{\includegraphics[scale=0.2]{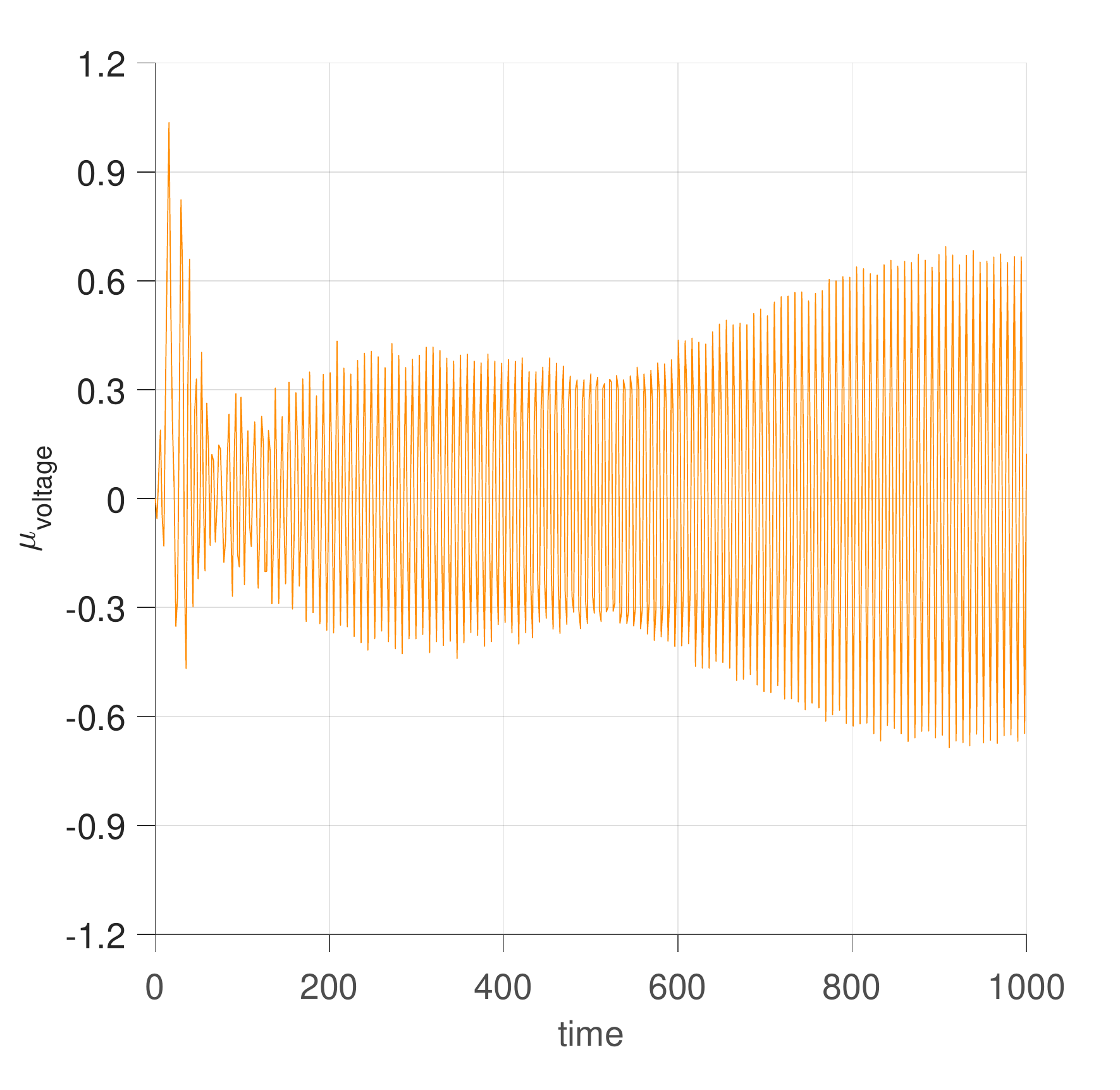}}
~\\
\subfigure[$\tau_{corr}/\tau_{\Omega} = 1\%$]{\includegraphics[scale=0.2]{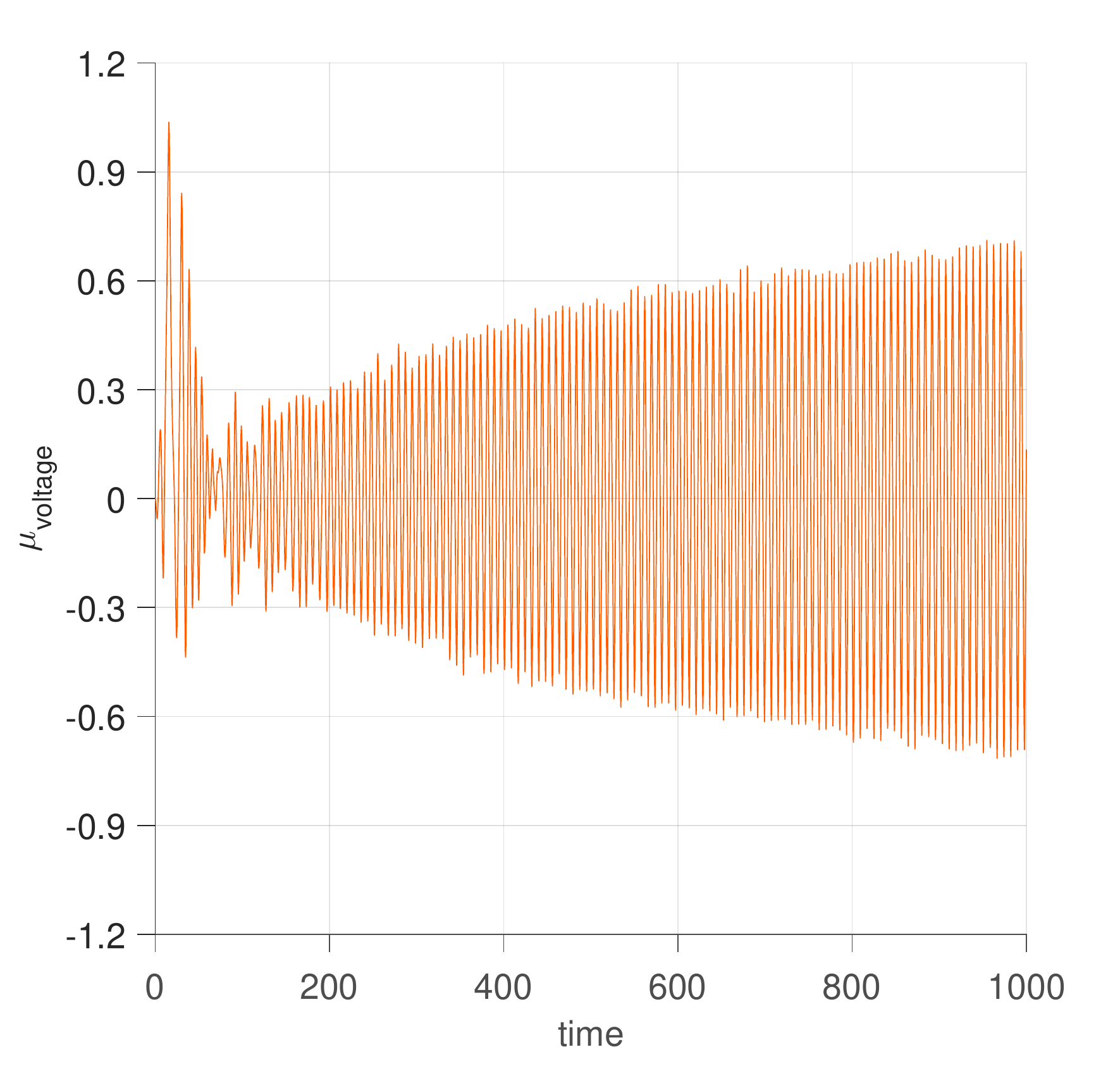}}
\subfigure[$\tau_{corr}/\tau_{\Omega} = 50\%$]{\includegraphics[scale=0.2]{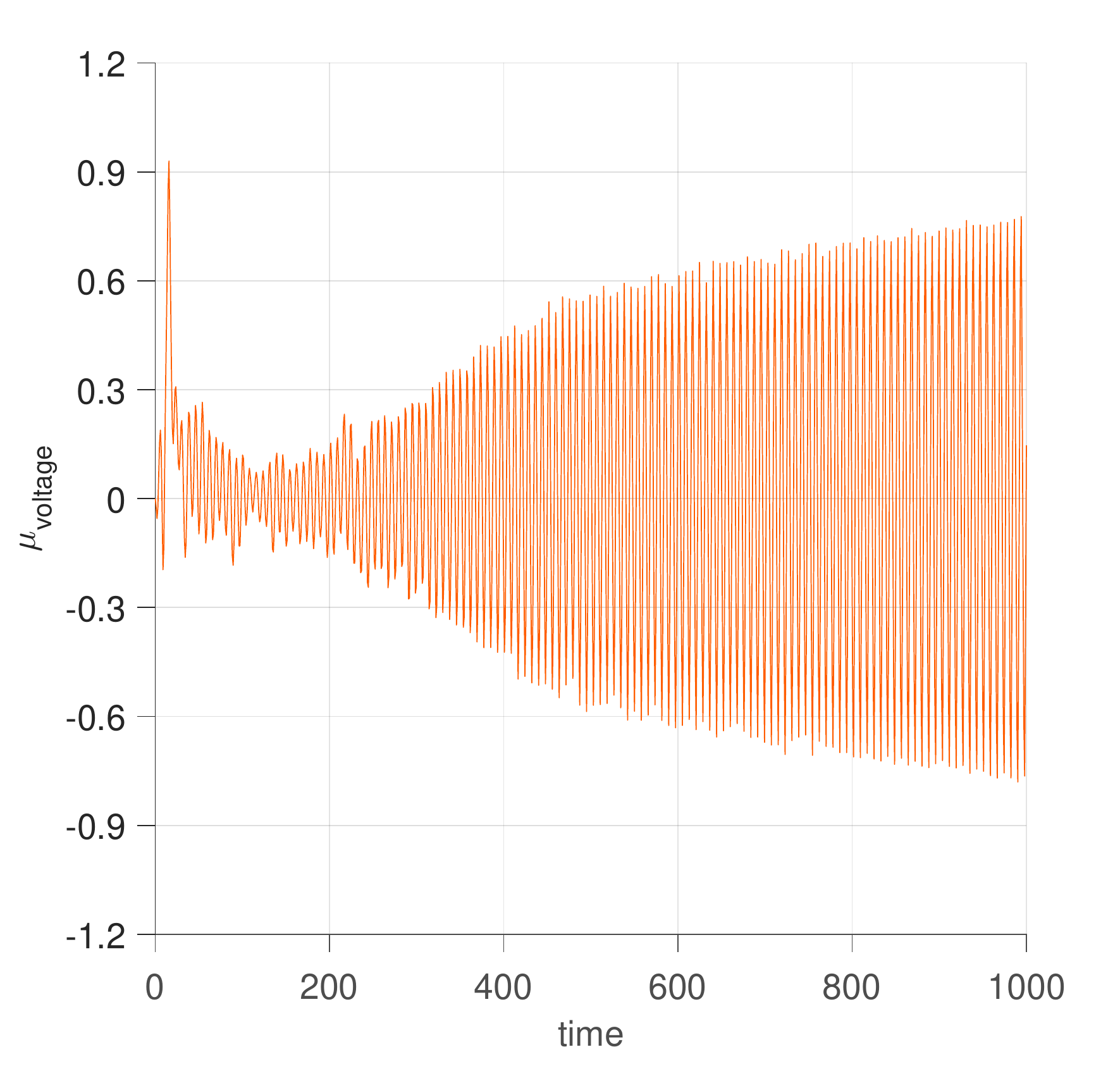}}
\subfigure[$\tau_{corr}/\tau_{\Omega} = 100\%$]{\includegraphics[scale=0.2]{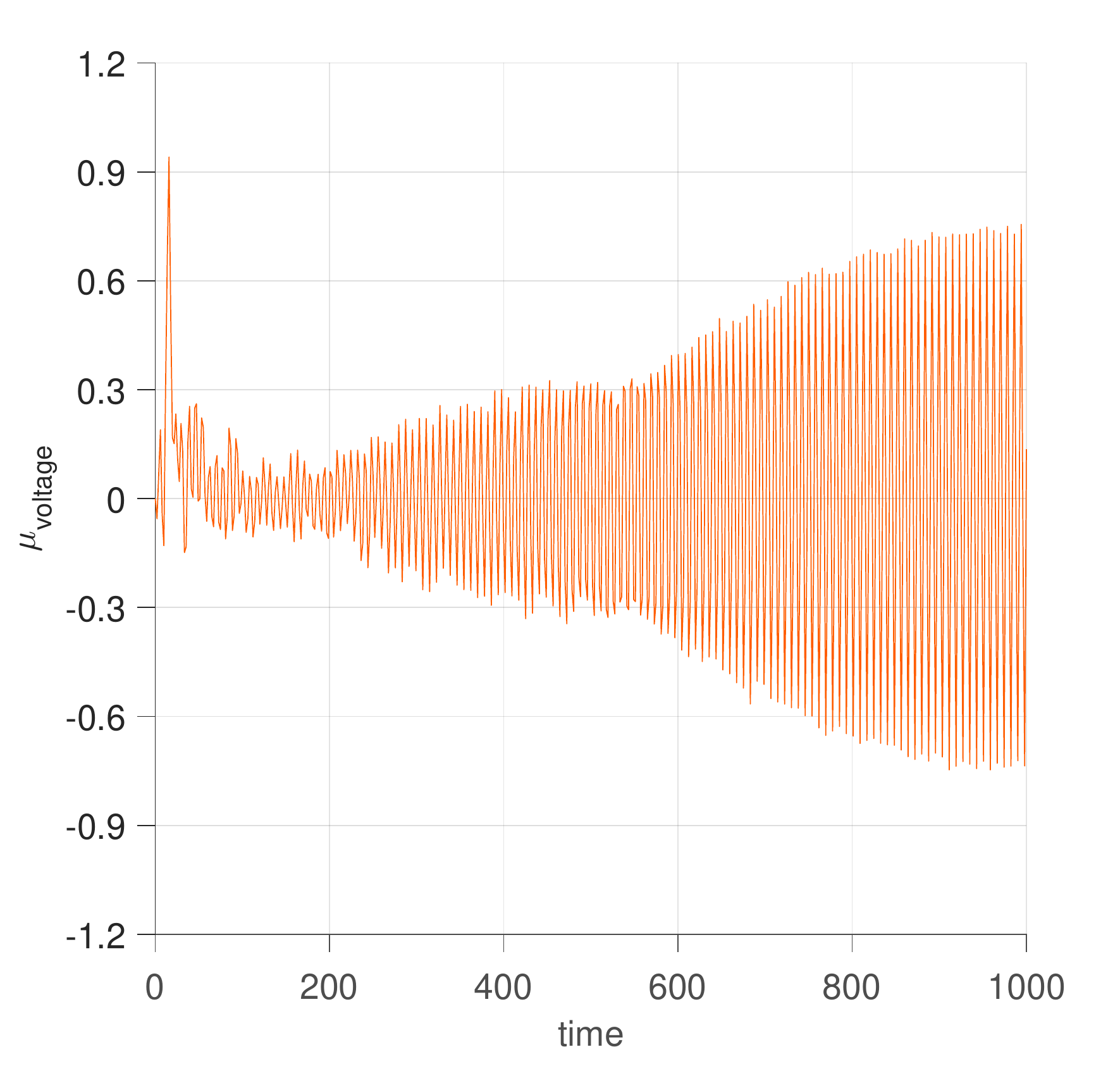}}
~\\
\subfigure[$\tau_{corr}/\tau_{\Omega} = 1\%$]{\includegraphics[scale=0.2]{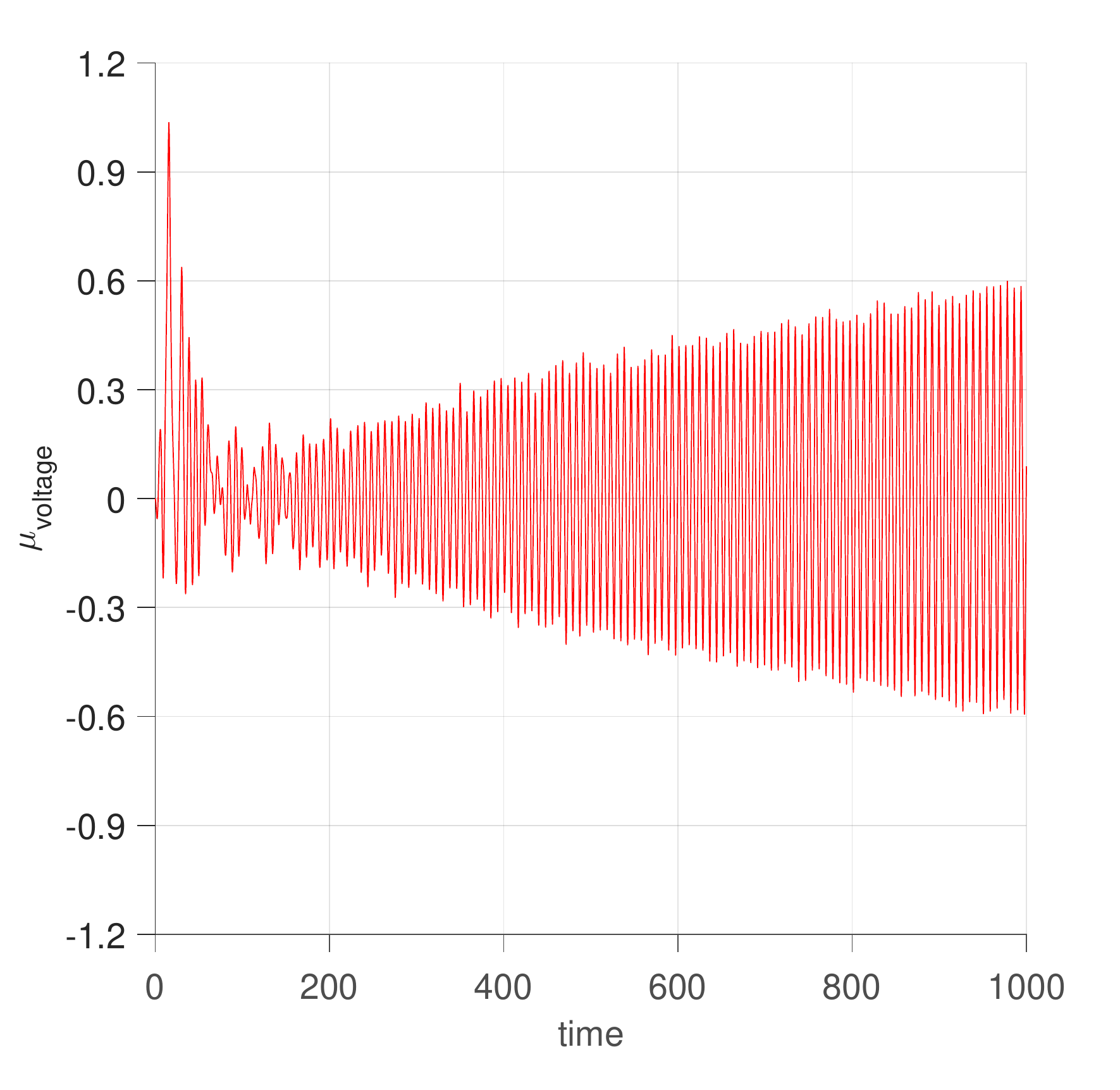}}
\subfigure[$\tau_{corr}/\tau_{\Omega} = 50\%$]{\includegraphics[scale=0.2]{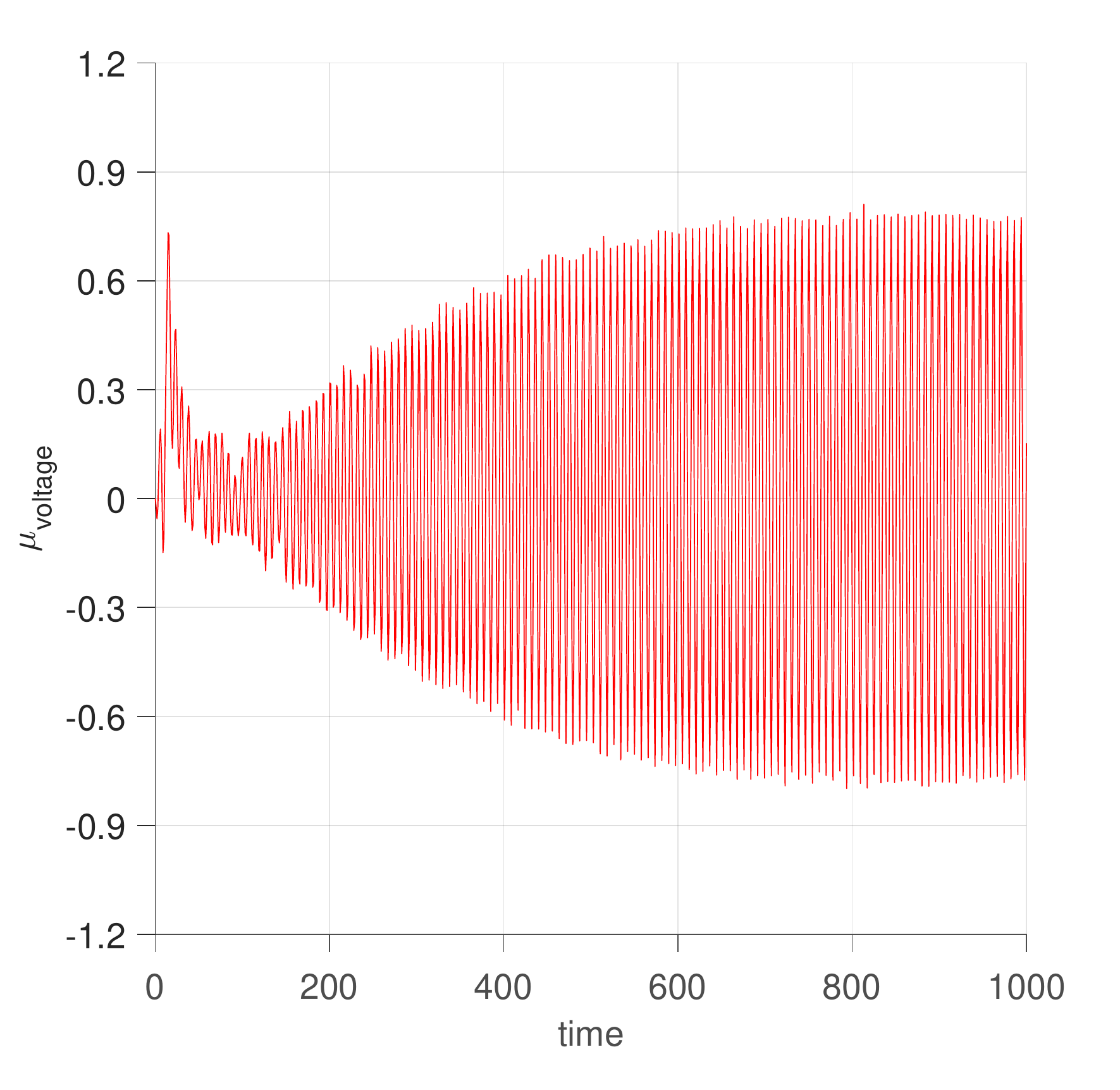}}
\subfigure[$\tau_{corr}/\tau_{\Omega} = 100\%$]{\includegraphics[scale=0.2]{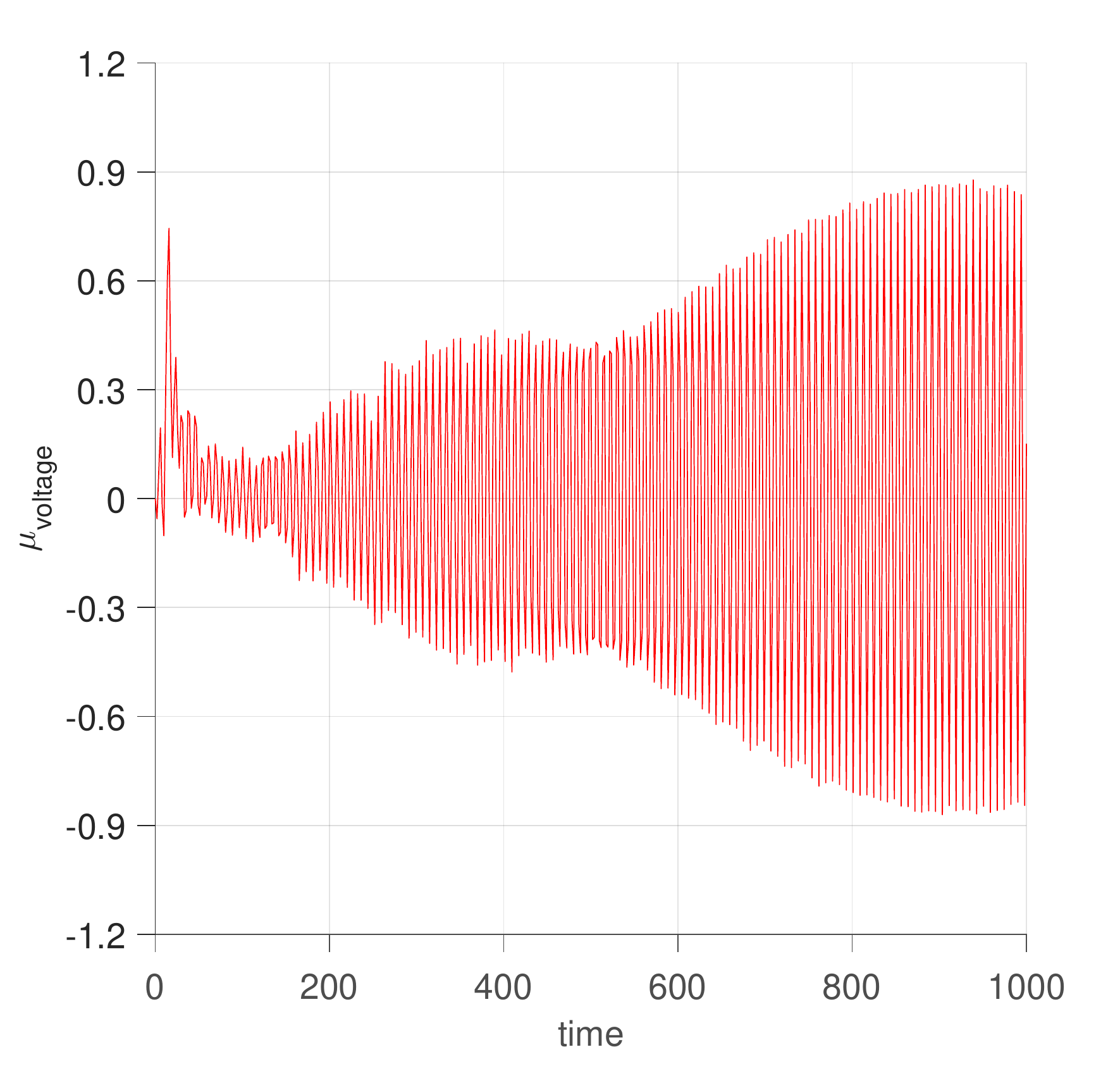}}

\caption{Voltage time series mean for $\sigma/f = 1\%$ (top), $\sigma/f = 25\%$ (middle) 
and $\sigma/f = 50\%$ (bottom), and different values of correlation time.}
\label{fig_mean}
\end{figure}

\begin{figure}[ht!]
\centering 
\subfigure[$\tau_{corr}/\tau_{\Omega} = 1\%$]{\includegraphics[scale=0.2]{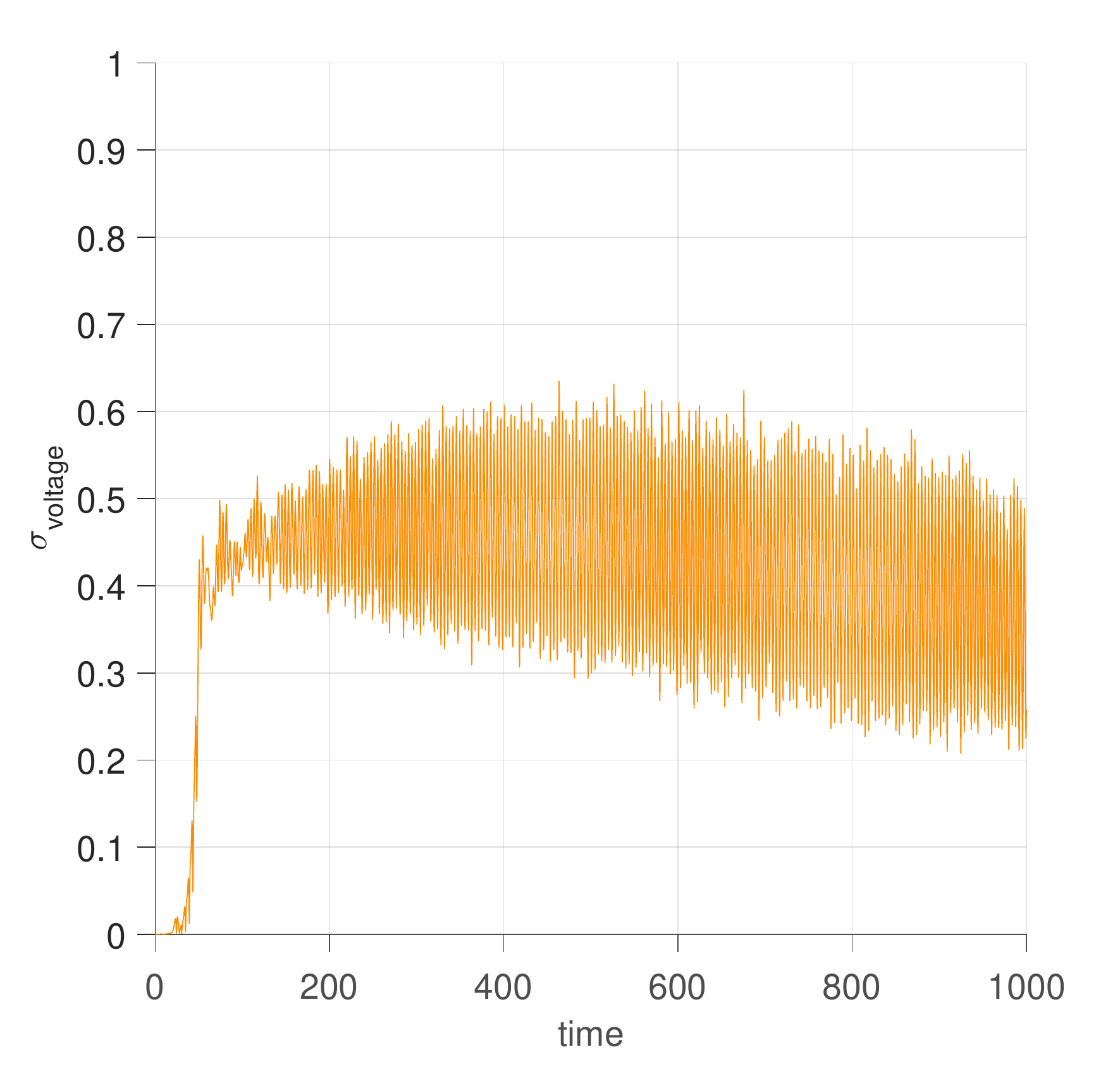}} 
\subfigure[$\tau_{corr}/\tau_{\Omega} = 50\%$]{\includegraphics[scale=0.2]{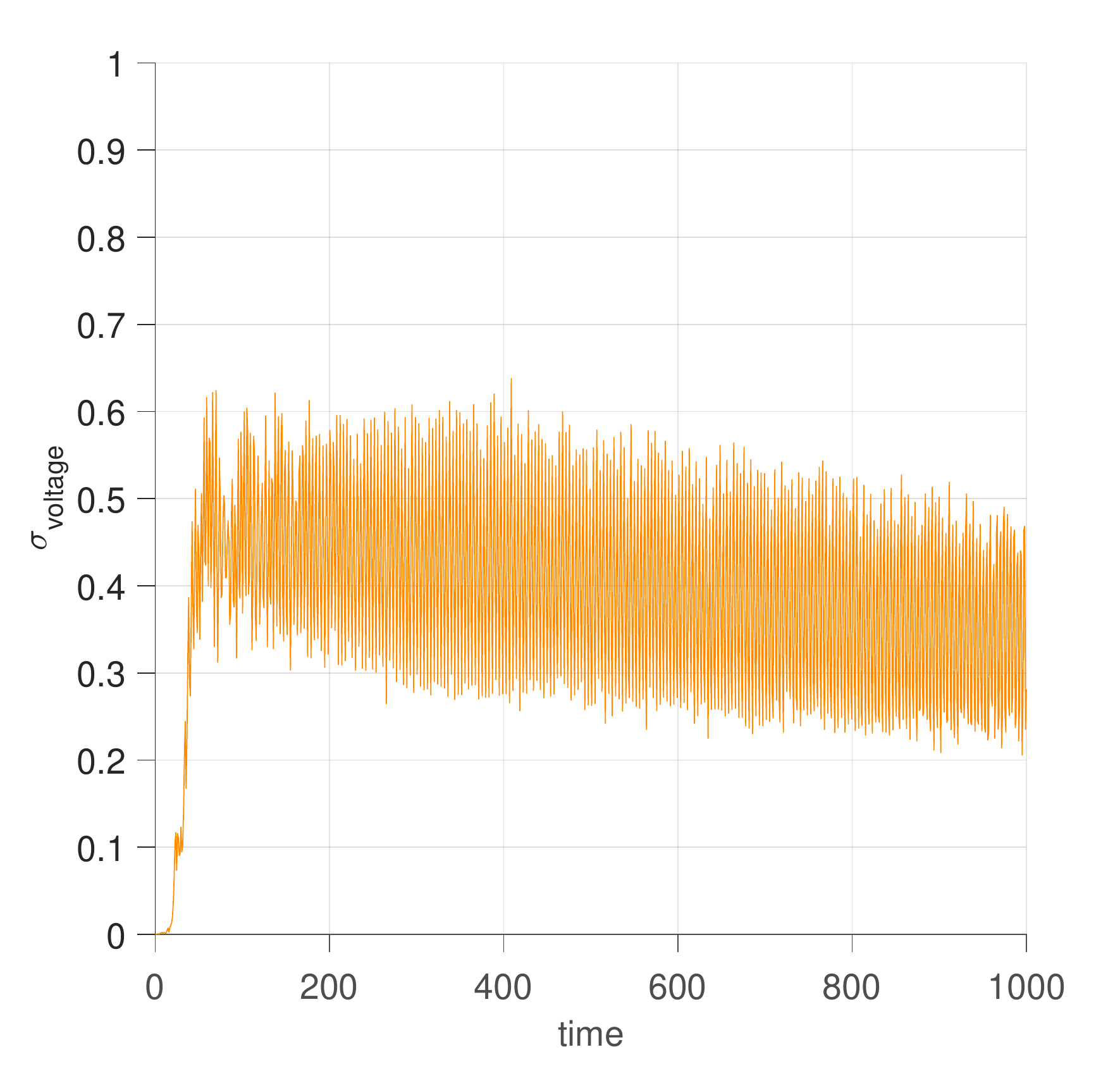}} 
\subfigure[$\tau_{corr}/\tau_{\Omega} = 100\%$]{\includegraphics[scale=0.2]{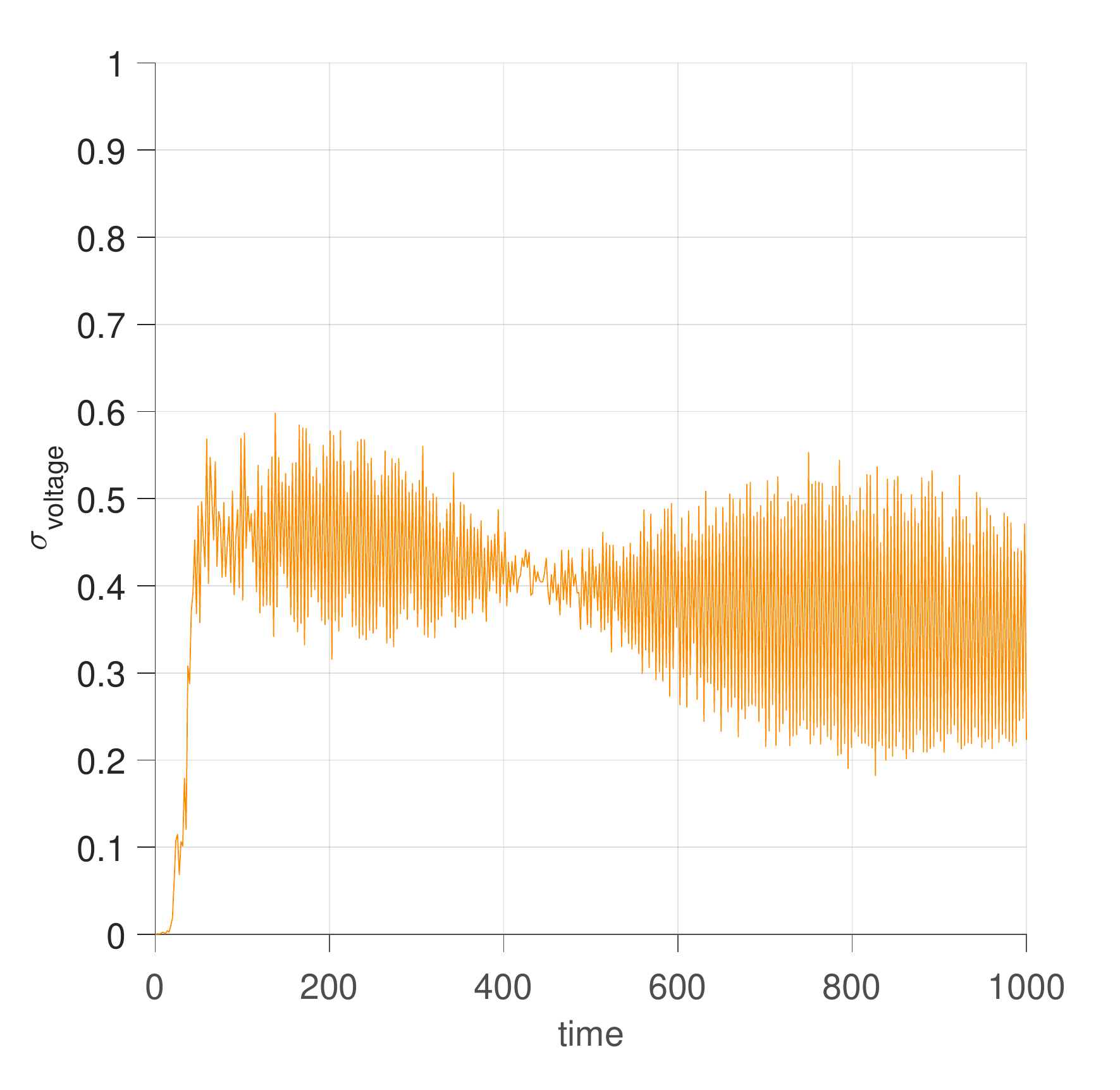}} 
~\\
\subfigure[$\tau_{corr}/\tau_{\Omega} = 1\%$]{\includegraphics[scale=0.2]{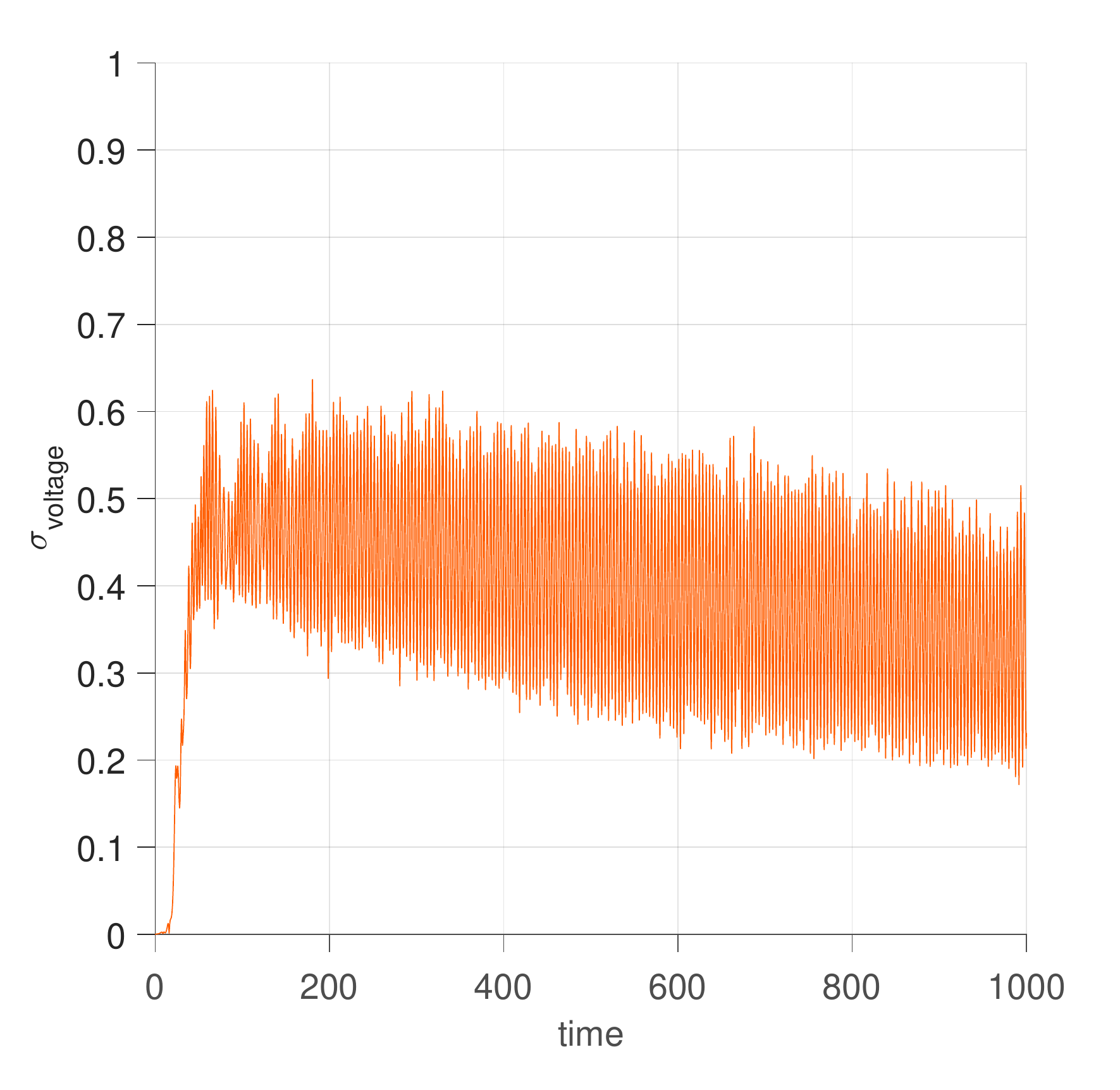}} 
\subfigure[$\tau_{corr}/\tau_{\Omega} = 50\%$]{\includegraphics[scale=0.2]{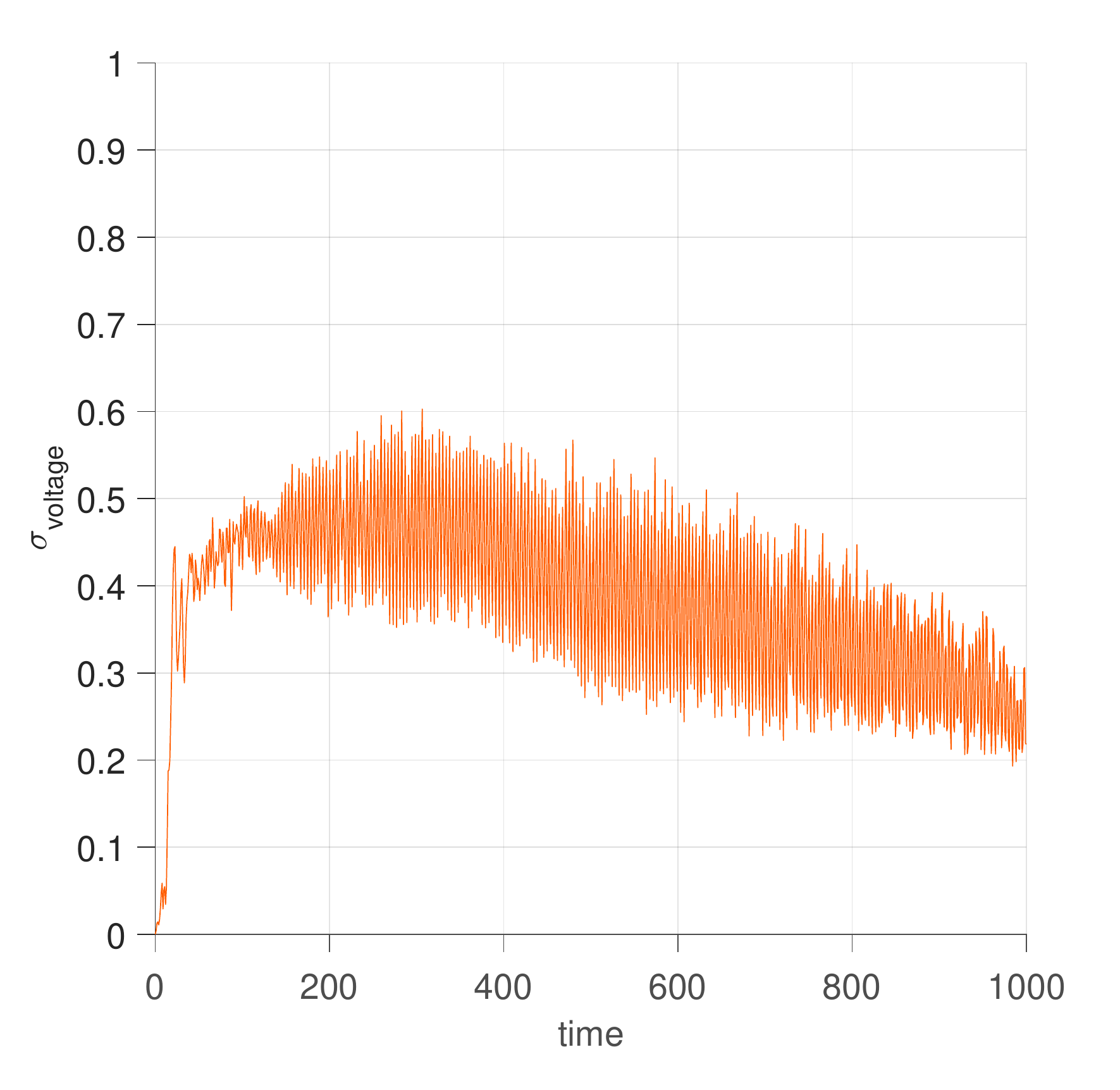}} 
\subfigure[$\tau_{corr}/\tau_{\Omega} = 100\%$]{\includegraphics[scale=0.2]{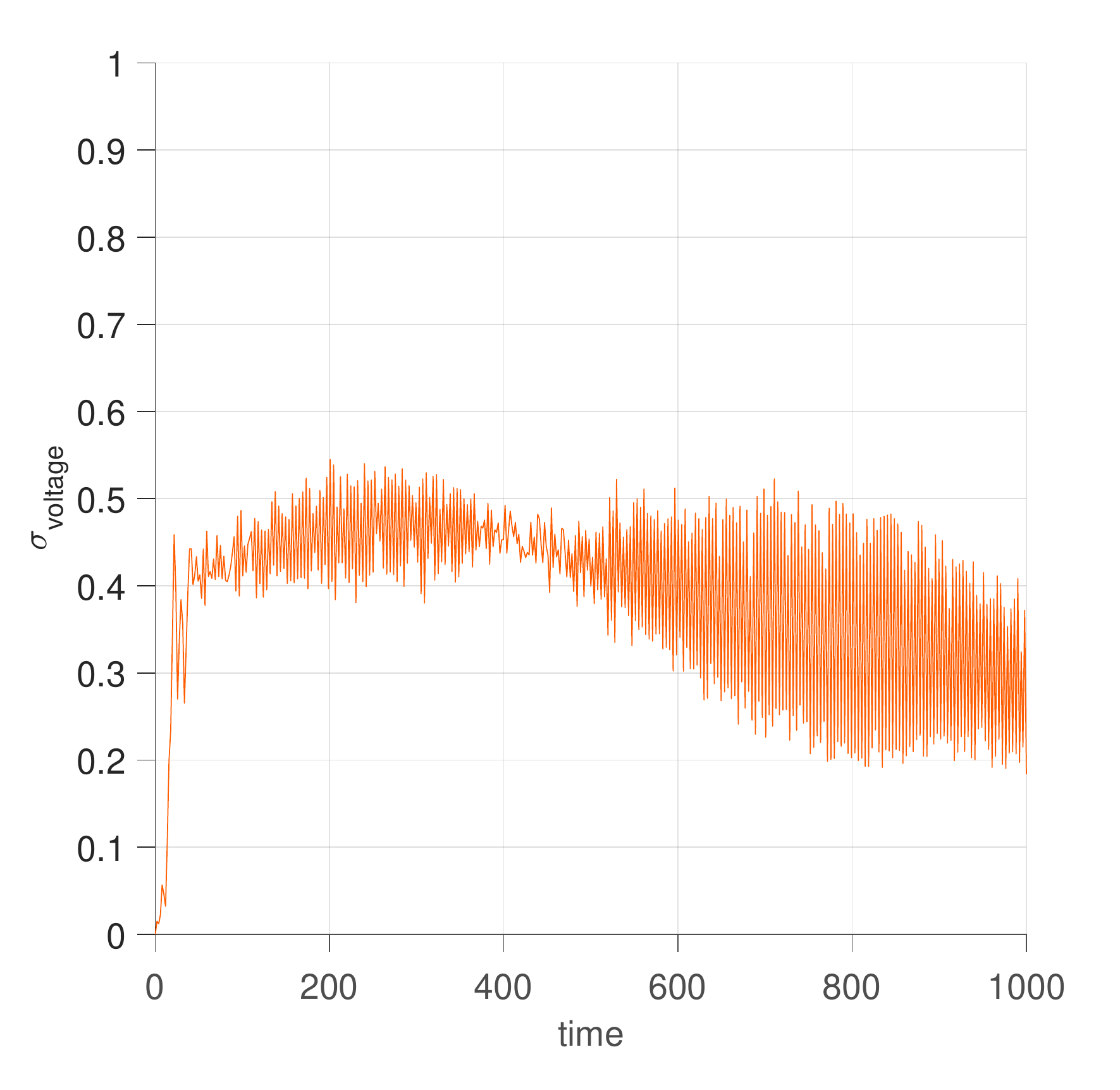}} 
~\\
\subfigure[$\tau_{corr}/\tau_{\Omega} = 1\%$]{\includegraphics[scale=0.2]{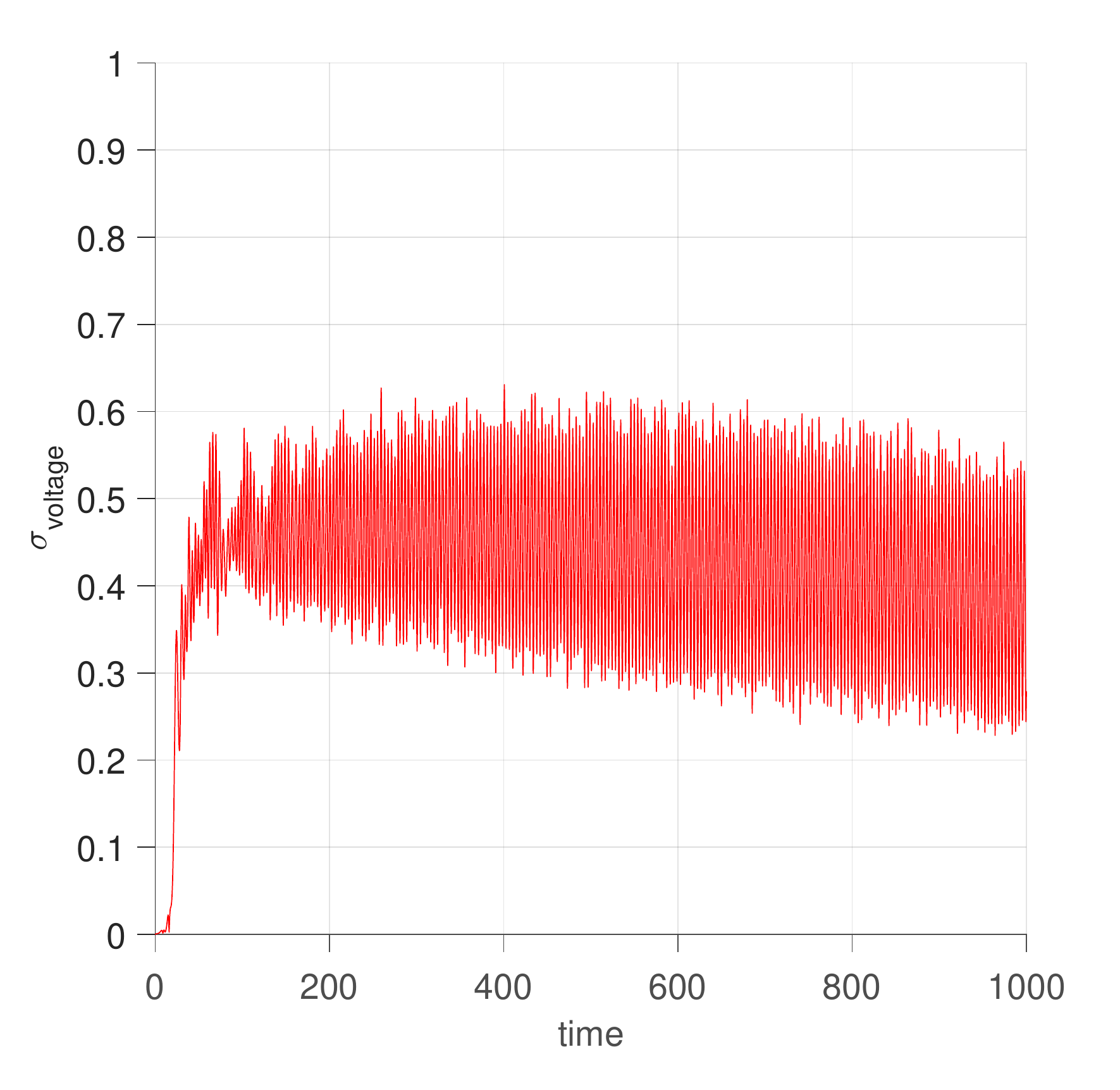}} 
\subfigure[$\tau_{corr}/\tau_{\Omega} = 50\%$]{\includegraphics[scale=0.2]{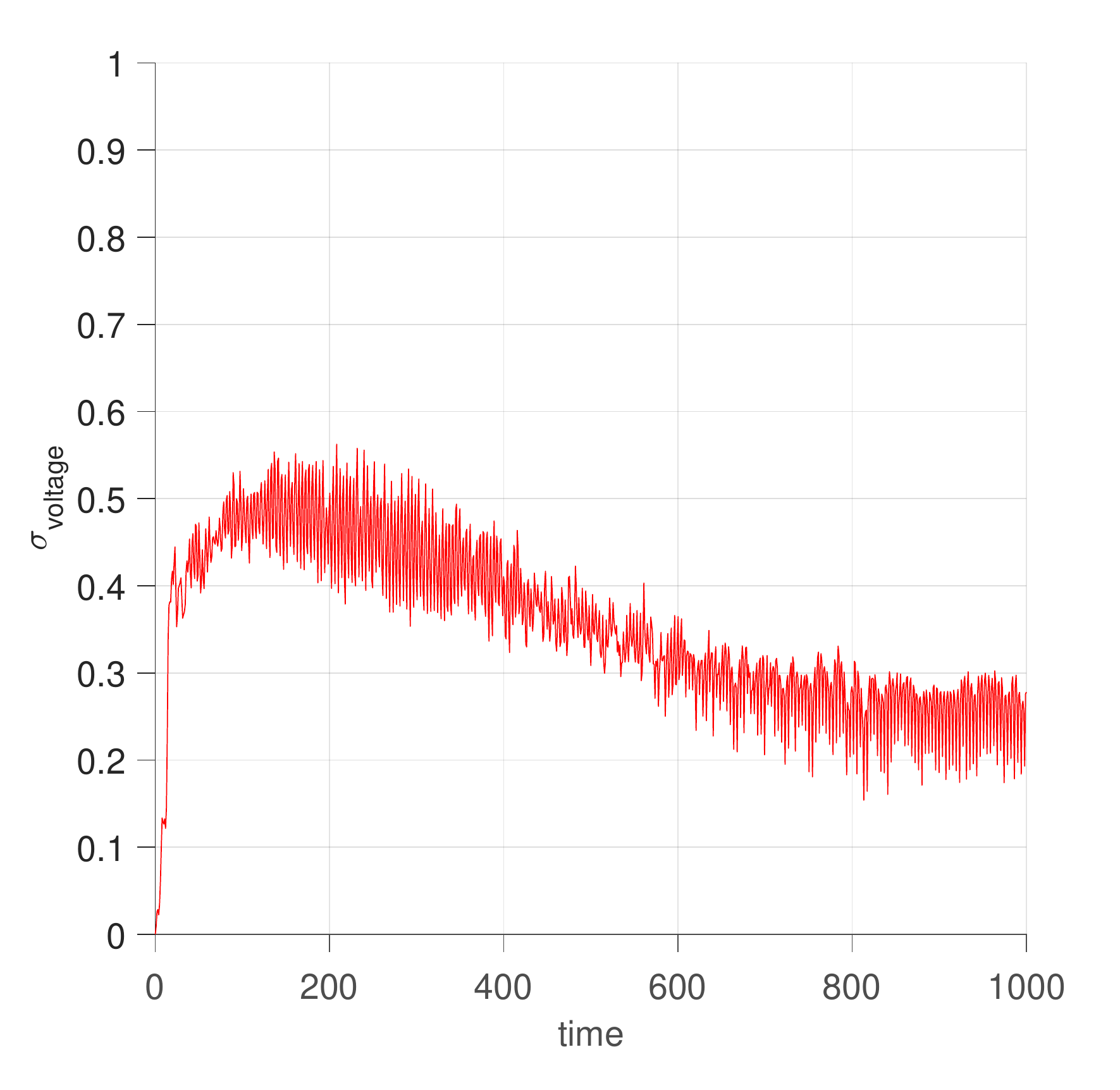}} 
\subfigure[$\tau_{corr}/\tau_{\Omega} = 100\%$]{\includegraphics[scale=0.2]{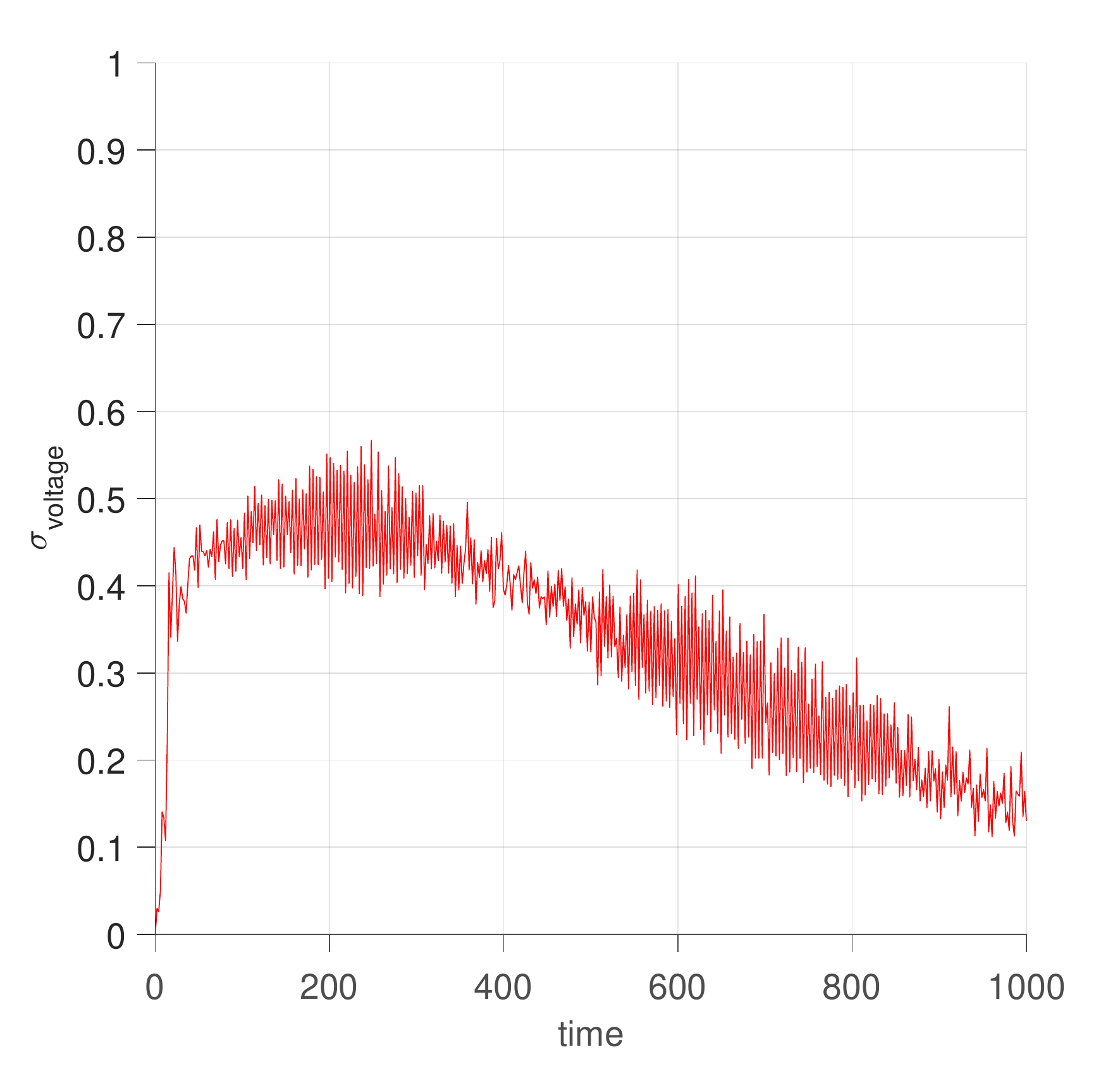}}

\caption{Voltage time series standard deviation for $\sigma/f = 1\%$ (top), $\sigma/f = 25\%$ (middle) 
and $\sigma/f = 50\%$ (bottom), and different values of correlation time.}
\label{fig_std}
\end{figure}

Figure \ref{fig_time_series_deltaN} presents different voltage time
series sampled from Monte Carlo simulations, 
for $\sigma/f = 1\%$, $\sigma/f = 25\%$ and $\sigma/f = 50\%$,
and several values of correlation times. Depicted on blue, the 
reader see the voltage response for a pure harmonic forcing. 
The corresponding mean and standard deviation values, for the 
are presented on Figures~\ref{fig_mean} and \ref{fig_std}, respectively.

The analysis of the Monte Carlo mean for the harvester output voltages 
on Figure~\ref{fig_mean} suggests a longer chaotic transient 
for the higher correlation times, specially for the maximum noise incidence, on
$\tau_{corr}/\tau_{\Omega} = 100\%$. In this case, the distortion observed 
in the voltage mean curve points the presence of more non-periodic time 
series during analyzed interval, despite of low level of noise.
As in the previous cases, the results analysis for higher
correlation times, with $\sigma/f = 25\%$, suggests longer transient 
states on voltage dynamics, what could be inferred trough the distortions 
on mean voltage signal. Although, the standard deviation curves points 
to less variations on output voltages regards the mean behavior, 
pointing to a typical pattern. 
A comparison between the results with $\tau_{corr}/\tau_{\Omega} = 1\%$
suggests that low correlation times reduce the effects of a higher noise 
incidence over the harmonic forcing.
For $\sigma/f = 50\%$, while the mean output voltage behavior follows 
those on the previous cases, the standard deviation curves for higher correlation 
times suggests less variations between the mean curve shape and a typical 
voltage time series.
The comparison between the standard deviation results for 
different levels of noise reinforces the correlation time influence
over the voltage. But note that, even when noise standard deviation
is high, a low correlation time suppress its effects over system 
voltage response. 

\quad

\pagebreak

\textbf{4. Final Remarks}
\label{sec_final_remarks}
\smallskip

The paper analyzed the influence of a random noise component on 
external harmonic forcing acting over a piezoelectric energy 
harvesting system. Numerical simulations considered different values 
noise correlation times and standard deviations, from which noise 
incidence and magnitude were tested. The results suggests a strong 
influence of noise signal over the voltage results for higher correlation 
times and standard deviations, what implies in longer chaotic transient 
states. Lower correlation times reduces the impacts of the higher noise 
magnitudes. For future works, authors intent to 
carry out more detailed analysis by expanding the noise correlation 
time and coefficient of variation and observation windows.

\quad

{\bf Acknowledgments}. 
The authors acknowledge the support 
given to this research by the funding agencies 
Carlos Chagas Filho Research Foundation of 
Rio de Janeiro State (FAPERJ) under grants 
E-26/010.002.178/2015 and E-26/010.000.805/2018, 
and Coordenação de Aperfeiçoamento de Pessoal 
de Nível Superior - Brasil (CAPES) - 
Finance Code 001.

\quad




\begin{thebibliography}{43}

\bibitem{Cottone2009}
F. Cottone, H. Vocca and L. Gammaitoni,
Nonlinear energy harvesting,
Phys. Rev. Lett., 102:080601, 2009.
https://doi.org/10.1103/PhysRevLett.102.080601


\bibitem{Bradai2015}
S. Bradai, S. Naifar, C. Viehweger, O. Kanoun and G. Litak,
Nonlinear analysis of an electrodynamic broadband energy harvester,
The European Physical Journal Special Topics, 224:2919--2927, 2015.
https://doi.org/10.1140/epjst/e2015-02598-0


\bibitem{Kamalinejad2015}
P. Kamalinejad, C. Mahapatra, Z. Sheng, S. Mirabbasi, V. C. M. Leung and Y. L. Guan,
Wireless energy harvesting for the Internet of Things,
IEEE Communications Magazine, 53:102-108, 2015.
https://doi.org/10.1109/MCOM.2015.7120024

\bibitem{Vijayan2015}
K. Vijayan, M.I. Friswell, H. Haddad Khodaparast, S. Adhikari,
Non-linear energy harvesting from coupled impacting beams,
International Journal of Mechanical Sciences, 96-97:101--109, 2015.
https://doi.org/10.1016/j.ijmecsci.2015.03.001


\bibitem{Abdelkareem2018}
M. A. A. Abdelkareem, L. Xu, M. K. A. Ali, A. Elagouz, J. Mi, S. Guo, Y. Liu and L. Zuo,
Vibration energy harvesting in automotive suspension system: A detailed review,
Applied Energy, 229:672--699, 2018.
https://doi.org/10.1016/j.apenergy.2018.08.030

\bibitem{Wang2018}
C. Wang, Q. Zhang, W. Wang and J. Feng,
A low-frequency, wideband quad-stable energy harvester using combined nonlinearity and frequency up-conversion by cantilever-surface contact,
Mechanical Systems and Signal Processing, 112:305--318, 2018.
https://doi.org/10.1016/j.ymssp.2018.04.027


\bibitem{Cunhajr_matec2018}
J. V. L. L. Peterson, V. G. Lopes and A. {Cunha~Jr},
Dynamic analysis and characterization of a nonlinear bi-stable piezo-magneto-elastic energy harvester,
MATEC Web of Conferences, 241:01001, 2018.
https://doi.org/10.1051/matecconf/201824101001


\bibitem{Borowiec2015}
M. Borowiec,
Energy harvesting of cantilever beam system with linear and nonlinear piezoelectric model,
The European Physical Journal Special Topics, 224:2771--2785, 2015.
https://doi.org/10.1140/epjst/e2015-02588-2


\bibitem{Langley2015}
R. S. Langley,
Bounds on the vibrational energy that can be harvested from random base motion,
Journal of Sound and Vibration, 339:247--261, 2015.
https://doi.org/10.1016/j.jsv.2014.11.012


\bibitem{Erturk2009}
A. Erturk, J. Hoffmann and D. J. Inman,
A piezomagnetoelastic structure for broadband vibration energy harvesting,
Applied Physics Letters, 94:254102, 2009.
https://doi.org/10.1063/1.3159815

\end{thebibliography}
\end{document}